\documentclass[a4paper,11pt]{article}
\usepackage[a4paper, margin=.11\paperwidth]{geometry}

\usepackage[export]{adjustbox} 
\usepackage{amsbsy}
\usepackage{amsmath}
\usepackage{amsopn}
\usepackage{amssymb}
\usepackage{amsthm} 
\usepackage[toc,page]{appendix}
\usepackage{array} 
\usepackage{booktabs}
\usepackage{bm} 
\usepackage{braket} 
\usepackage{cancel} 
\usepackage{calc} 
\usepackage{capt-of} 
\usepackage{caption} 
\usepackage{cite}
\usepackage{comment} 
\usepackage{dcolumn} 
\usepackage{diagbox} 
\usepackage{enumerate} 
\usepackage{esint} 
\usepackage{extarrows} 
\usepackage{footmisc} 
\usepackage{footnote} 
\usepackage{graphicx}
\usepackage{booktabs}
\usepackage[colorlinks=true, linkcolor=blue, citecolor=red, urlcolor=magenta]{hyperref}
\usepackage{ifpdf} 
\usepackage{gensymb}
\usepackage{listings}
\usepackage{makecell}
\usepackage{makeidx}
\usepackage{mathrsfs}
\usepackage[version=3]{mhchem} 
\usepackage{multicol}
\usepackage{multirow} 
\usepackage[section]{placeins}
\usepackage[displaymath,textmath,sections,graphics,floats]{preview}
\usepackage{pythonhighlight}
\usepackage{setspace}
\usepackage{sidecap} 
\usepackage{simplewick} 
\usepackage{siunitx} 
\usepackage{slashed}
\usepackage{soul}
\usepackage{subcaption}
\usepackage{threeparttable} 
\usepackage{tcolorbox} 
\usepackage{upgreek}
\usepackage{url}
\usepackage{xcolor}

\allowdisplaybreaks

\begin{document}
\begin{titlepage}
\begin{center}
{\bf\Large
\boldmath{
Exact parametrization of a minimal seesaw model}
}
\\[12mm]
Zi-Qiang~Chen,$^{1,2,3}$\footnote{E-mail: \texttt{chenziqiang22@mails.ucas.ac.cn}}
Xi-He~Hu,$^{1,2,3}$\footnote{E-mail: \texttt{huxihe23@mails.ucas.ac.cn}}
and
Ye-Ling~Zhou$^{1}$\footnote{E-mail: \texttt{zhouyeling@ucas.ac.cn}}
\\[-2mm]
\end{center}
\vspace*{0.50cm}
\begin{center}
$^{1}$\it School of Fundamental Physics and Mathematical Sciences,\\
  	\it Hangzhou Institute for Advanced
	Study, UCAS, Hangzhou 310024, China\\
	$^{2}$\it Institute of Theoretical Physics, Chinese Academy of Sciences, Beijing 100190, China\\
	$^{3}$\it University of Chinese Academy of Sciences, Beijing 100049, China
\end{center}

\vspace*{1.20cm}

\begin{abstract}
{\indent
We propose a parametrization of neutrino masses and mixing in the minimal seesaw model (MSM). The MSM, which introduces two heavy sterile neutrinos, is the minimal extension of the Standard Model in addressing the tiny masses of active neutrinos. The parametrization includes 11 free parameters: 6 neutrino oscillation parameters (2 mass-squared differences $\Delta m^2_{21}$, $\Delta m^2_{31}$, 3 mixing angles $\theta_{12}$, $\theta_{13}$, $\theta_{23}$, and 1 Dirac phase $\delta_{\rm CP}$), 1 mass parameter in $0\nu2\beta$ decay $m_{ee}$, and 4 additional parameters: 2 heavy neutrino masses $M_1$ and $M_2$, 1 active-sterile mixing angle $\theta_{14}$ and 1 CP-violating phase $\delta_{14}$. This parametrization is derived exactly from the most general neutrino mass matrix in the MSM without any approximation. We further discuss its implications in phenomenological studies.
}
\end{abstract}
\end{titlepage}

\section{Introduction}

Neutrino masses and mixing, as proven in neutrino oscillation experiments, are definitely new physics beyond the Standard Model (SM) \cite{ParticleDataGroup:2024cfk}. A straightforward way to understand these tiny masses is provided by the canonical seesaw mechanism \cite{Minkowski:1977sc, Yanagida:1979as, Gell-Mann:1979vob, Glashow:1979nm, Mohapatra:1979ia}: the smallness of light neutrino masses is suppressed by large masses of heavy mediators beyond the SM. We address the mechanism with $n$ copies of right-handed (RH) sterile neutrinos $N_{\rm R}$ into account. Relevant Lagrangian terms for neutrino masses are listed to be
\begin{align} 
-{\cal L}_\nu = \overline{\ell_{\rm L}} Y_\nu \widetilde{H} N_{\rm R} + \frac12 \overline{(N_{\rm R})^c} M_{\rm R} N_{\rm R} + {\rm h.c.} \,,
\end{align}
where $\ell_{\rm L}$ denotes the three copies of SM lepton doublets, $\widetilde{H} = {\rm i}\sigma_2 H^*$ with the SM Higgs doublet $H$, $Y_\nu$ is a $3\times n$ Yukawa coupling matrix and $M_{\rm R}$ is a $n\times n$ Majorana mass matrix for $n$ copies of RH neutrinos. After the electroweak symmetry breaking, a $(3+n) \times (3+n)$ Majorana mass matrix is obtained in the basis $\big((\nu_{\rm L})^c, N_{\rm R}\big)$, 
\begin{align}
{\cal M} = \begin{pmatrix} 0 & M_{\rm D} \\ M_{\rm D}^T & M_{\rm R} \end{pmatrix} \,,
\end{align}
where $M_{\rm D} = Y_\nu \langle H \rangle$. This matrix is diagonalized via an unitary matrix ${\cal U}$ as
\begin{align} \label{eq:U}
{\cal M} = 
{\cal U}  \begin{pmatrix} D_\nu & 0 \\ 0 & D_N \end{pmatrix} {\cal U}^T \,,\quad 
{\cal U}=
\begin{pmatrix} U & R \\ \tilde{S} & V \end{pmatrix}  \,.
\end{align}
The first block of ${\cal M}$ gives the {\it exact seesaw formula} \cite{Xing:2011ur}, 
\begin{align} \label{eq:exact_ss}
&U D_\nu U^T =  - R D_N R^T\,.
\end{align}
In the limit of large hierarchy between the Dirac mass matrix and Mojorana mass matrix $M_{\rm D} \ll M_{\rm R}$, it is convenient to use the approximations $R \approx M_{\rm D}^{} M_{\rm R}^{-1} V$ and $D_N \approx V^{-1} M_{\rm R} (V^{-1})^T$. Then we arrive at the well-known {\it approximate seesaw formula}
\begin{align} \label{eq:approx_ss}
M_\nu \equiv U D_\nu U^T \approx -M_{\rm D}^{} M_{\rm R}^{-1} M_{\rm D}^T \,.
\end{align}
Although it is approximate, it provides a simple explanation to connect the tiny active neutrino masses with the heavy particles: While the $M_D$ cannot be higher than the electroweak scale, the heavier Majorana masses for RH neutrinos lead to the smaller Majorana masses for light neutrinos. 

Adding RH neutrinos not only constitutes the simplest possibility for the origin of neutrino masses and mixing, 
but also supplies sources of peculiar signatures of experiments in particle physics, such as charged lepton flavor violations (cLFVs) \cite{Bilenky:1977du,Lee:1977qz,Marciano:1977wx,Cheng:1980tp} and collider searches for heavy neutral leptons \cite{delAguila:2007qnc,Atre:2009rg}.
On the other hand, the Majorana nature of RH neutrinos and the potential CP-violating source in the Dirac Yukawa interactions provide a well-known mechanism, baryogenesis via leptogenesis \cite{Fukugita:1986hr}, to explain the matter-antimatter asymmetry as observed in our universe \cite{Planck:2018vyg, ParticleDataGroup:2024cfk}. 
The approximate seesaw formula provides a clear vision on these phenomenological studies. 

The flavor space of light and heavy neutrinos involves a lot of independent parameters and a parametrization is necessary in  phenomenological studies. 
A widely used method is the Casas-Ibarra parametrization \cite{Casas:2001sr} based on the approximate seesaw formula.
It is originally introduced in parametrizing the cLFV processes \cite{Casas:2001sr} and then turns out to be very useful in the study of leptogenesis \cite{Davidson:2002qv}. In the minimal case, only two RH neutrinos are necessary to be introduced \cite{King:1999mb, King:2002nf, Frampton:2002qc}. 
This so-called the minimal seesaw model (MSM) provides the most economical and renormalizable framework to explain two non-zero neutrino mass-squared differences and the lepton flavor mixing observed in neutrino oscillation experiments. The MSM can effectively reduce free parameters in the flavor space. Several parametrizations specifically for the MSM have been proposed and were summarized in an early review \cite{Guo:2006qa}. For example, the natural reconstruction parametrization \cite{Barger:2003gt}, derived also from the approximate seesaw formula, provides a direct insight into the Dirac mass matrix in terms of the light neutrino mass matrix.

In this work, we will propose a new parametrization in the MSM framework. It is derived directly from the exact seesaw formula in Eq.~\eqref{eq:exact_ss} without any approximation. In the rest of this paper, we show the derivation of the Exact parametrization of the MSM in section~\ref{sec:EMSM}. The comparisons with Casas-Ibarra, Natural Reconstruction and other parametrizations are given in section~\ref{sec:comparepara}. The implications of this parametrization in phenomenological studies including experimental searches and leptogenesis are discussed in section~\ref{sec:pheimp}. We conclude in section~\ref{sec:conclusion}.

\section{The EMSM parametrization} \label{sec:EMSM}
The minimal seesaw model (MSM) introduces two right-handed (RH) sterile neutrinos $(N_{\rm R})_i\ (i=1, 2)$. 
The mass terms in the Lagrangian are expressed  in the $\nu-N_{\rm R}$ basis,
\begin{align}
   -\mathcal{L_{\rm MSM}}  \supset \frac{1}{2}\ \,
    \overline{
    \begin{pmatrix}
    \nu_{\rm L} & (N_{\rm R})^c 
    \end{pmatrix}}
    \begin{pmatrix}
    0 & M_{\rm D}^{}\\
    M_{\rm D}^{T} & M_{\rm R}^{}
    \end{pmatrix}
    \begin{pmatrix}
    (\nu_{\rm L})^c \\ N_{\rm R}^{}
    \end{pmatrix}+{\rm h.c.} \,,
\end{align}
where $M_{\rm R}$ is a $2\times 2$ Majorana mass matrix of RH neutrinos, $M_{\rm D}$ is a $3\times 2$ Dirac mass matrix after the Higgs get the vacuum expectation value $\braket{H}=246/\sqrt{2}$~GeV, and $Y_{\nu}$ is a Dirac Yukawa coupling.
The full $5\times 5$ neutrino matrix composed of $M_{\rm D}$ and $M_{\rm R}$ are zero determinate. Thus, the lightest neutrino mass must be zero. So far, the mass hierarchy of three light neutrinos has not been determined.
There are two options, in the normal hierarchy (NH), we have $0=m_1<m_2<m_3\ll M_1\le M_2$ in the MSM and in the inverted hierarchy (IH), it is $0=m_3<m_1<m_2\ll M_1\le M_2$, where we have denoted the indices for the two heavy neutrinos following their mass ordering.
The well-known seesaw formula for light neutrino masses in Eq.~\eqref{eq:approx_ss}
is obtained via 1st-order approximation under the condition $M_{\rm D} \ll M_{\rm R}$. In this work, we will abandon this approximation and adopt the exact seesaw formula in Eq.~\eqref{eq:exact_ss}. This exact seesaw relation has already been suggested in the study of neutrinoless double beta ($0\nu 2 \beta$) decay analysis \cite{Fang:2021jfv}. 

With only two RH neutrinos under this consideration, the full $5\times 5$ neutrino mass matrix ${\cal M}$ can be orthogonally diagonalized by a $5\times 5$ unitary matrix $\mathcal{U}$,
\begin{align} 
	\begin{pmatrix}
		0 & M_{\rm D}\\
		M_{\rm D}^{T} & M_{\rm R}
		\end{pmatrix} =
		\mathcal{U} \begin{pmatrix}
		D_{\nu} & 0\\
		0 & D_{N}
		\end{pmatrix} \mathcal{U}^{T} \,,
		\label{eq:diagonalization}
	\end{align}
where $D_\nu = {\rm diag}\{m_1, m_2, m_3\}$ and $D_{\rm N} = {\rm diag}\{M_1, M_2\}$ are light and heavy neutrino masses,  respectively.
Here, $\mathcal{U}$ can be written as  \cite{Xing:2007zj,Xing:2011ur}
\begin{align}
    \mathcal{U}=
    \begin{pmatrix}
    I & 0\\
    0 & U_0'
    \end{pmatrix}&
    \begin{pmatrix}
    A & R\\
    S & B
    \end{pmatrix}
    \begin{pmatrix}
    U_0 & 0\\
    0 & I
    \end{pmatrix}\,, \label{eq:mathU}
\end{align}
where $U_0$ and $U_0'$ are $3 \times 3$ and $2 \times 2$ unitary matrix, respectively. 
$A$, $R$, $S$ and $B$ are $3 \times 3$, $3 \times 2$, $2 \times 3$, $2 \times 2$ matrices, respectively.
Due to the unitarity property of $\mathcal{U}$, they satisfy
\begin{align}
	\begin{aligned}
		&AA^{\dagger}+RR^{\dagger}=BB^{\dagger}+SS^{\dagger}=I \,, \\ 
		&AS^{\dagger}+RB^{\dagger}=A^{\dagger}R+S^{\dagger}B=0 \,,\\ 
		&A^{\dagger}A+S^{\dagger}S=B^{\dagger}B+R^{\dagger}R=I \,.
	\end{aligned}\label{equ:nonunitarity}
\end{align}
Comparing Eq.~\eqref{eq:mathU} with Eq.~\eqref{eq:U}, we have correlations $U = A U_0$, $\tilde{S} = U_0' S U_0$ and $V = U_0' B$.
Meanwhile, the unitary matrix $\mathcal{U}$ can be expressed by the Euler-like parametrization with 10 two-dimensional unitary matrices $O_{ij}$ $(1\leq i<j \leq 5)$,
\begin{align}
	\begin{pmatrix}
		U_0 & 0 \\
		 0  & I
	\end{pmatrix}
	=O_{23}^{}O_{13}^{}O_{12}^{}, \quad \begin{pmatrix}
		I & 0 \\
		 0  & U_0'
	\end{pmatrix}&
	=O_{45}^{}, \quad 
	\begin{pmatrix}
		A & R \\
		 S  & B
	\end{pmatrix}
	=O_{35}^{}O_{25}^{}O_{15}^{}O_{34}^{}O_{24}^{}O_{14}^{} 
	\label{eq:Oij} \,.
\end{align}
Here, $O_{ij}$ consists of $c_{ij}\equiv\cos{\theta}_{ij}$ and $\hat{s}_{ij}\equiv e^{i\delta_{ij}}\sin{\theta}_{ij}$.
$\theta_{ij}$ is the rotation angle in the $(i,j)$ plane. $\theta_{12}$, $\theta_{13}$ and $\theta_{23}$ denote the active-active mixing, $\theta_{14}$, $\theta_{15}$, $\theta_{24}$, $\theta_{25}$, $\theta_{34}$ and $\theta_{35}$ represent the active-sterile mixing and $\theta_{45}$ is the sterile-sterile mixing angle.
$\delta_{ij}$ is the corresponding complex phase along with $\theta_{ij}$. 

The exact expression of $U^{}_0$ is
\begin{align}
U^{}_0 = \begin{pmatrix} c^{}_{12} c^{}_{13} & \hat{s}^*_{12}
c^{}_{13} & \hat{s}^*_{13} \cr
-\hat{s}^{}_{12} c^{}_{23} -
c^{}_{12} \hat{s}^{}_{13} \hat{s}^*_{23} & c^{}_{12} c^{}_{23} -
\hat{s}^*_{12} \hat{s}^{}_{13} \hat{s}^*_{23} & c^{}_{13}
\hat{s}^*_{23} \cr
\hat{s}^{}_{12} \hat{s}^{}_{23} - c^{}_{12}
\hat{s}^{}_{13} c^{}_{23} & -c^{}_{12} \hat{s}^{}_{23} -
\hat{s}^*_{12} \hat{s}^{}_{13} c^{}_{23} & c^{}_{13} c^{}_{23}
\cr \end{pmatrix} \,.
\label{eq:U0}
\end{align}
It is emphasized that this $U_0$ is different from the PMNS matrix.
In the standard parametrization, the PMNS matrix $V_0$ is
\begin{align}
	V^{}_0 =
	\begin{pmatrix} c^{}_{12} c^{}_{13} & s_{12}
	c^{}_{13} & s_{13}e^{-i\delta_{\rm CP}} \cr
	-s^{}_{12} c^{}_{23} -
	c^{}_{12} s^{}_{13} s_{23} e^{i\delta_{\rm CP}} & c^{}_{12} c^{}_{23} -
	s_{12} s^{}_{13} s_{23}e^{i\delta_{\rm CP}} & c^{}_{13}
	s_{23} \cr
	s^{}_{12} s^{}_{23} - c^{}_{12}
	s^{}_{13} c^{}_{23}e^{i\delta_{\rm CP}} & -c^{}_{12} s^{}_{23} -
	s^{}_{12} s^{}_{13} c^{}_{23}e^{i\delta_{\rm CP}} & c^{}_{13} c^{}_{23} \cr
    \end{pmatrix}
	\begin{pmatrix}
		e^{i\rho} & 0 & 0\cr
		0 & e^{i\sigma} & 0 \cr
		0 & 0 & 1
	\end{pmatrix} \, ,
	\label{eq:PMNS}
	\end{align}
where $\delta_{\rm CP}$ is the Dirac CP-violating phase and $\rho$ and $\sigma$ are two Majorana phases.
In the MSM, the phase $\rho$ is unphysical and taken as $\rho=0$.
We can correlate $U_0$ in the Euler-like parametrization to $V_0$ in the standard parametrization, through an orthogonal transformation of the light neutrino mass basis.
Hence, the relations between $\delta_{ij}$ of $U_0$ and $\delta_{\rm CP},\rho,\sigma$ of $V_0$ are $\delta_{\rm CP}=\delta_{13}-\delta_{12}-\delta_{23}$, $\rho = \delta_{12}+\delta_{23}=0$ and $\sigma = -\delta_{12}$.
The exact expressions of $A$, $R$, $S$ and $B$ are
\begin{align}
	&\begin{aligned}
		A=
		\begin{pmatrix} c^{}_{14} c^{}_{15}  & 0 & 0
			\cr \vspace{-0.45cm} \cr
			\begin{array}{l}    -c^{}_{14} \hat{s}^{}_{15} \hat{s}^*_{25} 
				-\hat{s}^{}_{14} \hat{s}^*_{24} c^{}_{25}
			\end{array} &
				c^{}_{24} c^{}_{25}  & 0 \cr \vspace{-0.45cm} \cr
			\begin{array}{l}  - c^{}_{14} \hat{s}^{}_{15} c^{}_{25} \hat{s}^*_{35} 
				-\hat{s}^{}_{14} c^{}_{24} \hat{s}^{*}_{34}c^{}_{35}  \\+ \hat{s}^{}_{14} \hat{s}^*_{24} \hat{s}^{}_{25} \hat{s}^*_{35}
			\end{array} &
			\begin{array}{l}
				\hspace{0.04cm}   -c^{}_{24} \hat{s}^{}_{25} \hat{s}^*_{35}  -\hat{s}^{}_{24} \hat{s}^*_{34} c^{}_{35} 
			\end{array} &
				c^{}_{34} c^{}_{35}  \cr 
			\end{pmatrix}  \; ,
	\end{aligned} \label{eq:A}\\
	&\begin{aligned}
		R = \begin{pmatrix} \hat{s}^*_{14} c^{}_{15}  &
			\hat{s}^*_{15}  \cr \vspace{-0.45cm} \cr
			\begin{array}{l}  -\hat{s}^*_{14} \hat{s}^{}_{15} \hat{s}^*_{25}  
				+ c^{}_{14} \hat{s}^*_{24} c^{}_{25}
			\end{array} &  
			c^{}_{15} \hat{s}^*_{25}  \cr \vspace{-0.45cm} \cr
			\begin{array}{l}  - \hat{s}^*_{14} \hat{s}^{}_{15}
				c^{}_{25} \hat{s}^*_{35} 
				- c^{}_{14} \hat{s}^*_{24} \hat{s}^{}_{25} \hat{s}^*_{35} + c^{}_{14} c^{}_{24} \hat{s}^*_{34} c^{}_{35} 
			\end{array} &
			\begin{array}{l} 
				c^{}_{15} c^{}_{25} \hat{s}^*_{35}
			\end{array} \cr
		    \end{pmatrix}  \; ,
	\end{aligned} \label{eq:R} \\
	&\begin{aligned}
		S = \begin{pmatrix} -\hat{s}^{}_{14} c^{}_{24} c^{}_{34}  &
			-\hat{s}_{24}c_{34} &  \quad-\hat{s}_{34} \cr \vspace{-0.45cm} \cr
			\begin{array}{l}  \hat{s}^{}_{14}c_{24}\hat{s}^*_{34} \hat{s}^{}_{35}+\hat{s}_{14}^{}\hat{s}_{24}^{*}\hat{s}_{25}^{}c_{35}\\-c_{14}\hat{s}_{15}c_{25}^{}c_{35}^{}  
			\end{array} & 
			\hat{s}^{}_{24} \hat{s}^*_{34}\hat{s}^{}_{35} -c_{24}\hat{s}_{25}c_{35}& \quad -c_{34}\hat{s}_{35}  \cr \vspace{-0.45cm} 
		    \end{pmatrix} \; , 			
	\end{aligned} \label{eq:S} \\ 
	&\begin{aligned}
		B =\begin{pmatrix} c^{}_{14} c^{}_{24}c^{}_{34}  & 0 \cr \vspace{-0.45cm} \cr
			\begin{array}{l}    -c^{}_{14}c^{}_{24}\hat{s}^{*}_{34}\hat{s}^{}_{35}
				-c^{}_{14} \hat{s}^{*}_{24} \hat{s}^{}_{25} c^{}_{35}
				-\hat{s}^{*}_{14} \hat{s}^{}_{15} c^{}_{25} c^{}_{35}
			\end{array} & 
			\quad c^{}_{15} c^{}_{25} c^{}_{35}\cr \vspace{-0.45cm}  \cr \vspace{-0.45cm}
		   \end{pmatrix}   \; ,
	\end{aligned} \label{eq:B}
\end{align}
where $A$ is a non-unitarity matrix.
By setting $N_{\rm R}$ in the mass eigenstates, i.e. $D_{N}=M_{\rm R}$, we can fix $\theta_{45}=0$ and $U'_0=I$.

For convenience, one usually adopts $U = AU_0$ in Eq.~\eqref{eq:U} to describe the flavor mixing of three light neutrinos, and the effective mass matrix of light neutrinos is $M_\nu=UD_\nu U^T$ in the neutrino flavor basis.
For the following calculations, we take $\tilde{R} \equiv A^{-1} R$ as the active-sterile mixing matrix of neutrinos, 
\begin{align}
	\begin{aligned}
		\tilde{R}  =  
		\begin{pmatrix} 
		\displaystyle\frac{\hat{s}_{14}^*}{c_{14}}  \quad
		& \displaystyle \frac{\hat{s}_{15}^*}{c_{14}c_{15}}\\[4mm]
		\displaystyle \frac{\hat{s}_{24}^*}{c_{14}c_{24}}\quad
		&
		\displaystyle \frac{\hat{s}_{14}^{}\hat{s}_{15}^*\hat{s}_{24}^*}{c_{14}c_{15}c_{24}} +  \frac{\hat{s}_{25}^{*}}{c_{15}c_{24}c_{25}}
		\\[4mm]
		\displaystyle \frac{\hat{s}_{34}^*}{c_{14}c_{24}c_{34}}\quad & 
		\displaystyle \frac{\hat{s}_{14}^{}\hat{s}_{15}^{*}\hat{s}_{34}^{*}}{c_{14}c_{15}c_{24}c_{34}} +  
		\frac{\hat{s}_{24}^{}\hat{s}_{25}^*\hat{s}_{34}^*}{c_{15}c_{24}c_{25}c_{34}}+  
		 \frac{\hat{s}_{35}^{*}}{c_{15}c_{25}c_{34}c_{35}}
		\end{pmatrix} \,,
	\end{aligned} \label{eq:tildeR}
\end{align}
where $c_{ik}$ and $\hat{s}_{ik}$ are active-sterile mixing angles defined in Eq.~\eqref{eq:Oij}.
Actually, six active-sterile mixing angles and six phase angles are not independent, so we will solve Eq.~\eqref{eq:diagonalization} to find the correlation between $\theta_{ik}$ and $\delta_{ik}$ in the following.
From the upper left $3\times 3$ zero matrix in Eq.~\eqref{eq:diagonalization}, there is $U D_{\nu}U^{T}=-R D_NR^{T}$.
We define a new effective mass matrix of light neutrinos as
\begin{align}
	m_{ij}\equiv (U_0 D_\nu U_0^{T})_{ij} = -(\tilde{R} D_N \tilde{R}^{T})_{ij} \,, \label{eq:mij}
\end{align}
where $m_{ij}= \left(A^{-1}M_\nu(A^{-1})^T\right)_{ij}$.
From \eqref{eq:mij}, there are five independent equations, 
\begin{align}
	\begin{aligned}
		(\tilde{R} D_N \tilde{R}^{T})_{11} &= -m_{11} \,, \\
		(\tilde{R} D_N \tilde{R}^{T})_{22} &= -m_{22} \,, \\
		(\tilde{R} D_N \tilde{R}^{T})_{33} &= -m_{33} \,, \\
		w_1 \tilde{R}_{11} + w_2 \tilde{R}_{21} &+ w_3 \tilde{R}_{31} = 0 \,, \\
		w_1 \tilde{R}_{12} + w_2 \tilde{R}_{22} &+ w_3 \tilde{R}_{32} = 0 \,,
	\end{aligned}
\end{align}
where $w_i = (U_0^*)_{i1}$ for the NH and $w_i= (U_0^*)_{i3}$ for the IH with $i=1,2,3$.
Further, for the NH case, we solve the above five equations and obtain the following results \footnote{This involves the root operation for a complex number $z$, and we adopt $\sqrt{z}=\sqrt{|z|}\,\exp\{i\arg{(z)}/2\}$ and $\arg{(z)}\in(-\pi,\pi]$.},
\begin{align}
	\begin{aligned}
		\tilde{R}_{12} &= \eta_0 \sqrt{-\frac{m_{11}}{M_2} -\frac{M_1}{M_2} \tilde{R}_{11}^2} \,,\\
		\tilde{R}_{21} &= \frac{- (m_{11} w_1^2 + m_{22} w_2^{2} - m_{33} w_3^2)\, \tilde{R}_{11} +  2\, \eta_1 \sqrt{m_2m_3}\,W_1\, \tilde{R}_{12}}{2 m_{11} w_1 w_2} \,,\\
		\tilde{R}_{31} &= \frac{- (m_{11} w_1^2 - m_{22} w_2^{2} + m_{33} w_3^2)\, \tilde{R}_{11} -  2\, \eta_1 \sqrt{m_2m_3}\,W_1\, \tilde{R}_{12}}{2 m_{11} w_1 w_3} \,, \\
		\tilde{R}_{22} &= \frac{- (m_{11} w_1^2 + m_{22} w_2^{2} - m_{33} w_3^2)\, \tilde{R}_{12} -  2\, \eta_1 \sqrt{m_2m_3}\,W_2\, \tilde{R}_{11}}{2 m_{11} w_1 w_2} \,, \\
		\tilde{R}_{32} &= \frac{- (m_{11} w_1^2 - m_{22} w_2^{2} + m_{33} w_3^2)\, \tilde{R}_{12} +  2\, \eta_1 \sqrt{m_2m_3}\,W_2\, \tilde{R}_{11}}{2 m_{11} w_1 w_3} \,,
	\end{aligned}\label{eqn:Rij}
\end{align}
where $\eta_0,\eta_1=\pm 1$ are arbitrary signs and $W_i$ is an abbreviation as
\begin{align}
W_i=\frac{\sqrt{M_1M_2}}{M_i}\left[(U_0)_{23}(U_0)_{32}-(U_0)_{22}(U_0)_{33}\right](U_0^*)_{21}(U_0^*)_{31} \,.
\end{align}
We can see from the above formulas that complex $\tilde{R}_{11}$ is free and others of $\tilde{R}$ are decided by $\tilde{R}_{11}, M_1, M_2$ and low energy parameters.
From Eq.~\eqref{eq:tildeR}, $\tilde{R}_{11}$ is determined by the mixing angle $\theta_{14}$ and the corresponding phase $\delta_{14}$.
Thus, once $\theta_{14}$ and $\delta_{14}$ are given, we can obtain all the rest mixing angles and phases explicitly via Eq.~\eqref{eq:tildeR} and \eqref{eqn:Rij}.
Here, $\theta_{ik}$ and $\delta_{ik}$ are performed, 
\begin{align}
    \sin \theta_{ik} &= \frac{|p_{ik}|}{\sqrt{1+|p_{ik}|^2}} \,, \ \ \delta_{ik}= -\arg p_{ik} \,,
\end{align}
where $i=1,2,3$, $k=4,5$ and 
\begin{align}
    p_{24}&= c_{14} \tilde{R}_{21} \,,\label{eq:pa}\\
    p_{34}&= c_{14} c_{24} \tilde{R}_{31} \,,\label{eq:pb}\\
    p_{15}&= c_{14} \tilde{R}_{12} \,,\label{eq:pc}\\
    p_{25}&= c_{15}c_{24}\big(\tilde{R}_{22} - c_{14}^2 \tilde{R}_{11}^* \tilde{R}_{12} \tilde{R}_{21} \big) \,,\label{eq:pd}\\
    p_{35}&= c_{15}c_{25} c_{34}\big[\tilde{R}_{32} - c_{14}^2 c_{24}^2 ( \tilde{R}_{11}^* \tilde{R}_{12} + \tilde{R}_{21}^* \tilde{R}_{22} ) \tilde{R}_{31} \big] \,. \label{eq:pe}
\end{align}
In other words, given values of $\theta_{14}$ and $\delta_{14}$ (as well as RH neutrinos masses $M_1$ and $M_2$ and all low energy parameters used to determine $\tilde{R}_{ij}$), we solve out $\theta_{24}$ and $\delta_{24}$ via Eq.~\eqref{eq:pa}, and $\theta_{15}$ and $\delta_{15}$ via Eq.~\eqref{eq:pc}. Then, as $\theta_{24}$ and $\theta_{15}$ is known, we can obtain $\theta_{34}$ and $\delta_{34}$ via Eq.~\eqref{eq:pb}, and $\theta_{25}$ and $\delta_{25}$ via Eq.~\eqref{eq:pd}. Finally, $\theta_{35}$ and $\delta_{35}$ are obtained via Eq.~\eqref{eq:pe} after all other parameters are solved. 
Once all active-sterile mixing angles are derived, all submatrices $A$, $B$, $R$ and $S$ are obtained. 
From off-diagonal terms of Eq.~\eqref{eq:diagonalization}, the Dirac matrix $M_{\rm D}$ is exactly determined as
\begin{align}
M_{\rm D} = (AU_0) D_\nu (SU_0)^{T} + R D_N B^{T} \label{eq:MD} \,,
\end{align}
where $\theta_{45} = 0$ has been considered. 
Therefore, we parametrize the neutrino Dirac matrix $M_{\rm D}$ in the MSM by a new method.
Since we have not taken any approximation in the above derivation, we call this parametrization the {\it Exact parametrization of Minimal Seesaw Model} (EMSM). 

We summarize all free parameters in this parametrization,
\begin{align}
 m_2,\ m_3 (m_1),\ \theta_{12},\ \theta_{13},\ \theta_{23},\ \delta_{\rm CP},\ \sigma,\ M_1,\ M_2,\ \theta_{14},\ \delta_{14}\,,
\end{align}
for the NH (IH) case of light neutrino masses. Since the lightest neutrino mass is zero in the MSM, $m_2 = \sqrt{\Delta m^2_{21}}$, $m_3 = \sqrt{\Delta m^2_{31}}$ and $m_{ee} = |m_2 s_{12}^2 c_{13}^2 e^{2\sigma} + m_3 s_{13}^2 e^{-i 2\delta}|$ ($m_2 = \sqrt{\Delta m^2_{21}-\Delta m^2_{31}}$, $m_1 = \sqrt{-\Delta m^2_{31}}$ and $m_{ee} = |m_1 c_{12}^2 c_{13}^2 + m_2 s_{12}^2 c_{13}^2 e^{2\sigma}|$) for the NH (IH) in the MSM, where $m_{ee}$ is the effective mass parameter in $0\nu 2 \beta$ decay experiments.
We might choose another set of parameters,
\begin{align}
\Delta m^2_{21},\ \Delta m^2_{31},\ \theta_{12},\ \theta_{13},\ \theta_{23},\ \delta_{\rm CP},\ m_{ee},\ M_1,\ M_2,\ \theta_{14},\ \delta_{14}\,.
\end{align}
Here, the first 6 parameters are measured in neutrino oscillation experiments, the 7th is supposed to be measured in $0\nu 2 \beta$ experiments and the last 4 are parameters of sterile neutrinos. 
 
\section{Comparison between EMSM and other parametrizations} \label{sec:comparepara}
In this section, we will compare our parametrization (EMSM) with other parametrizations.
One well-known parametrization is the Casas-Ibarra (CI) parametrization, which is widely used in the study of neutrino phenomenology and cosmology. We compare the EMSM with the CI parametrization and, importantly, check the consistency between the two parametrizations.
We then turn to the comparison between the EMSM and the natural reconstruction (NR) parametrization. We show the similarity between them and point out the advantages of our parametrization. Comparisons with the rest parametrizations are summarized in the end of this section. 
Generally, most of these parametrizations adopt the approximate seesaw formula $M_\nu\simeq -M_{\rm D}^{}M_{\rm R}^{-1}M_{\rm D}^{T}$ under the condition $\sqrt{M_\nu/M_{\rm R}}\sim M_{\rm D}/M_{\rm R}\sim O(R)\ll 1$.
However, the EMSM uses the exact seesaw formula $M_\nu=-R D_N R^T$, so it is an exact parametrization, where the parameters in the EMSM have clear physical meanings.

\subsection{Casas-Ibarra parametrization} \label{sec:CIpara}
The CI parametrization adopts the approximate seesaw formula $M_{\nu}\simeq -M_{\rm D}^{}D_N^{-1}M_{\rm D}^{T}$ \cite{Casas:2001sr}. 
Applying this parametrization in the MSM \cite{Ibarra:2003up}, entries of the Dirac matrix $M_{\rm D}$ are approximately expressed as
\begin{align}
	\begin{aligned}
		(M_{\rm D}^{\rm CI})_{\alpha 1}^{} &\simeq +i\sqrt{M_1} \left[(V_0)_{\alpha 2} \sqrt{m_2^{}} \cos \omega + \xi\times (V_0)_{\alpha 3} \sqrt{m_3^{}} \sin \omega \right]\,, \\
		(M_{\rm D}^{\rm CI})_{\alpha 2}^{} &\simeq -i\sqrt{M_2} \left[(V_0)_{\alpha 2} \sqrt{m_2^{}} \sin \omega -\xi\times (V_0)_{\alpha 3} \sqrt{m_3^{}} \cos \omega \right] \,,
	\end{aligned}\label{eq:MDCI}
\end{align}
where $\alpha=1,2,3$ represents the charged lepton flavor index $\{e,\mu,\tau\}$, $\xi=\pm 1$ is an arbitrary sign, $\omega$ is a complex parameter, $V_0$ is the standard PMNS matrix as Eq.~\eqref{eq:PMNS} and more details are shown in Appendix~\ref{app:1}.

In the EMSM, the parametrization of $M_{\rm D}$ is shown on Eq.~\eqref{eq:MD}, which uses the exact seesaw formula.
When considering $D_\nu\ll D_N$, there is an approximation $M_{\rm D}^{\rm EMSM}\simeq R D_N B^T$.
Assuming that all active-sterile mixing angles are small enough, the submatrices of $\mathcal{U}$  approximate as $A_{\alpha i}\simeq \delta_{\alpha i}+O(\theta_{\alpha i}^2)$ and $B_{ij}\simeq \delta_{ij}+O(\theta_{ij}^2)$ via Eq.~\eqref{eq:A} and Eq.~\eqref{eq:B}.
Therefore, $M_{\rm D}^{\rm EMSM}$ in the leading-order approximation becomes  
\begin{align*}
	(M_{\rm D}^{\rm EMSM})_{\alpha i} & \simeq (R D_N)_{\alpha i}  \simeq M_i\, \theta_{\alpha i}\, e^{i\delta_{\alpha i}} \,,
\end{align*}
where the basis of $M_{\rm D}^{\rm EMSM}$ is chosen as the charged lepton flavor basis and the heavy neutrino mass basis same as $M_{\rm D}^{\rm CI}$ in Eq.~\eqref{eq:MDCI}.

\begin{figure}[h!]   
	\centering 
	\includegraphics[width=0.7\textwidth]{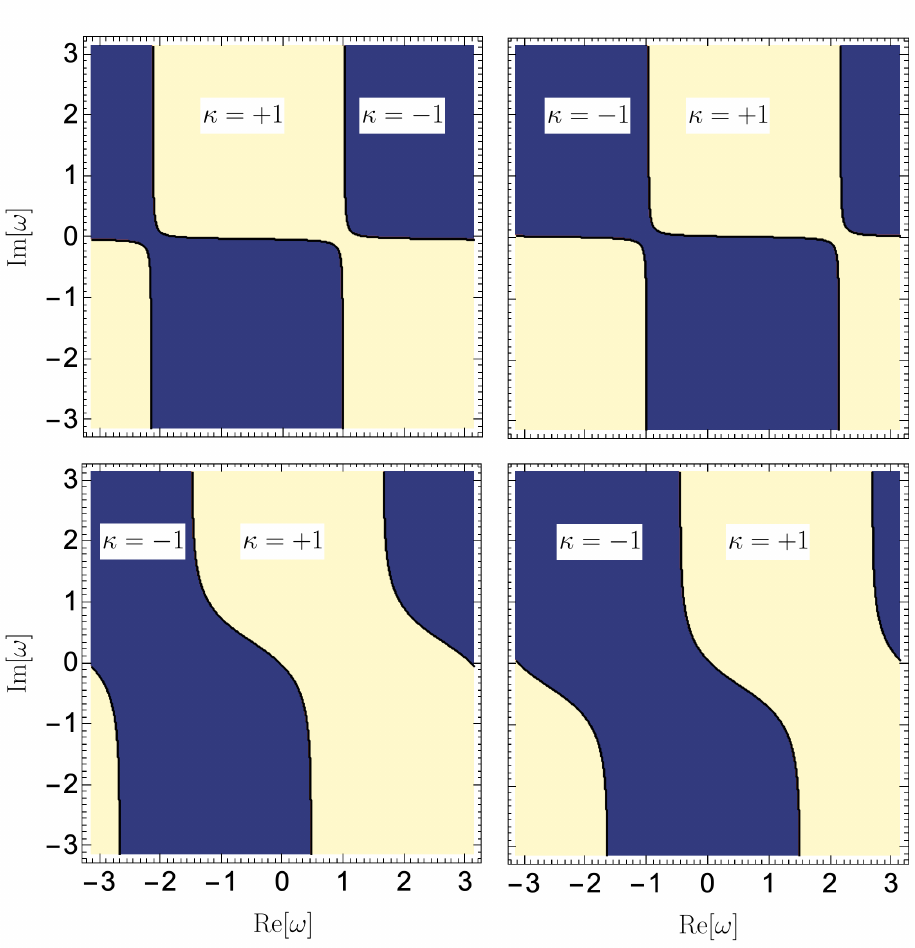}
	\caption{The $\kappa$ value about complex $\omega$ when $\xi=+1$ (left), $\xi=-1$ (right), and the Majorana phase $\sigma=0$ (upper), $\sigma=\pi/4$ (lower). The dark blue and bright yellow regions correspond to $\kappa=+1$ and $-1$, respectively.}
	\label{fig:kappa}  
\end{figure}

Sequently, we will correlate the EMSM with the CI parametrization, which is equivalent to find the formula of $\theta_{14},\delta_{14}$ and $\omega$.
Because of $(M_{\rm D}^{\rm EMSM})_{11}\simeq(M_{\rm D}^{\rm CI})_{11}$, we can directly obtain
\begin{align}
	\begin{cases}
		\theta_{14}\simeq|(V_0)_{12} \sqrt{m_2} \cos \omega + \xi\  (V_0)_{13} \sqrt{m_3} \sin \omega|/\sqrt{M_1}\\
		\delta_{14}\simeq-\arg[i((V_0)_{12} \sqrt{m_2} \cos \omega + \xi\  (V_0)_{13} \sqrt{m_3} \sin \omega)]	
	\end{cases} .\label{eq:equivalence}
\end{align}
Moreover, we must check the consistency of the rest terms between the EMSM and the CI parametrization.
From $(M_{\rm D}^{\rm EMSM})_{12}\simeq(M_{\rm D}^{\rm CI})_{12}$ and $(M_{\rm D}^{\rm EMSM})_{21}\simeq(M_{\rm D}^{\rm CI})_{21}$, there are two conditions $\eta_0=\kappa$ and $\eta_0\cdot\eta_1=\kappa\cdot\xi$, where $\kappa$ is defined as 
\begin{align}
	\kappa\equiv{\rm sign}\left\{\sqrt{(M_{D}^{\rm CI})_{12}^2}/(M_{D}^{\rm CI})_{12}\right\}=\pm 1 \,.
\end{align}
Here, $\kappa$ is independent of heavy neutrinos masses.
Finally, it is simple to check other terms $\{22,31,32\}$ satisfying $M_{\rm D}^{\rm EMSM}\simeq M_{\rm D}^{\rm CI}$.
We display the $\kappa$ value in Fig.~\ref{fig:kappa} when fixing the Majorana phase $\sigma=0$ and $\pi/4$.
Therefore, once $\xi$ is given and then $\kappa$ is obtained, the EMSM becomes consistent with the CI parametrization via Eq.~\eqref{eq:equivalence} and conditions
\begin{align}
	\kappa=\eta_0 \quad \text{and} \quad \xi=\eta_1 \,.
\end{align}

\subsection{Natural reconstruction} \label{sec:naturalrecon}
The natural reconstruction (NR) parametrization was proposed, based on the approximate seesaw formula and the MSM, to quantitatively reconstruct CP-violating phases for leptogenesis using only the neutrino low energy experiments \cite{Barger:2003gt}.
The Dirac matrix $M_{\rm D}^{\rm NR}$ can be obtained by the free parameter $(M_{\rm D}^{\rm NR})_{11}$, 
\begin{align}
	\begin{aligned}
		(M_{\rm D}^{\rm NR})_{12}& \simeq \zeta_1 \sqrt{-(M_{\nu})_{11}M_2-\frac{M_2}{M_1}(M_{\rm D}^{\rm NR})_{11}^2} \,, \\
		(M_{\rm D}^{\rm NR})_{\alpha 1} &\simeq \frac{1}{(M_\nu)_{11}} \left\{ (M_{\rm D}^{\rm NR})_{11} (M_\nu)_{1\alpha } + \zeta_\alpha (M_{\rm D}^{\rm NR})_{12} \sqrt{\frac{M_1}{M_2}} \sqrt{(M_\nu)_{11} (M_\nu)_{\alpha\alpha} - (M_\nu)_{1\alpha}^2} \right\} \, , \\
		(M_{\rm D}^{\rm NR})_{\alpha 2} &\simeq \frac{1}{(M_\nu)_{11}} \left\{ (M_{\rm D}^{\rm NR})_{12} (M_\nu)_{1\alpha } - \zeta_\alpha (M_{\rm D}^{\rm NR})_{11} \sqrt{\frac{M_2}{M_1}} \sqrt{(M_\nu)_{11} (M_\nu)_{\alpha\alpha } - (M_\nu)_{1\alpha}^2} \right\} \, ,	
	\end{aligned}
\end{align}
where $\alpha=2, 3$ represents the lepton flavor index $\{\mu,\tau\}$,  $\zeta_1=\pm 1$ is an arbitrary sign, $\det(M_\nu)=0$ constrains on $\zeta_\alpha=\pm 1$ and more details are shown in Appendix~\ref{app:1}.

The EMSM becomes consistent with the NR parametrization in the limit $M_{\rm D}\ll M_{\rm R}$, i.e., $M_{\rm D}^{\rm NR}\simeq M_{\rm D}^{\rm EMSM}\simeq RD_N$.
Here, for all corresponding elements of the both parametrizations are same, there are two conditions $\zeta_1=\eta_0$ and $\zeta_2=-\zeta_3=\eta_1$.
On the other hand, the EMSM is different from the NR parametrization.
The former is an exact parametrization while the latter is an approximate method. 
After getting the submatrix $\tilde{R}$ from Eq.~\eqref{eqn:Rij}, our parametrization can be used to calculate $\nu-N$ mixing angles $\theta_{ik}$ and corresponding phases $\delta_{ik}$ via Eqs.~\eqref{eq:pa}-\eqref{eq:pe} exactly.

\subsection{Other parametrizations} \label{sec:otherpara}
The Dirac mass matrix $M_{\rm D}$ of neutrinos is the key in the study of neutrino phenomenology, such as leptogenesis and charged lepton flavor violations, so parametrizing $M_{\rm D}$ is the foundation of some studies and many different parametrizations have been proposed.
These parametrizations, together with CI and NR, in the MSM framework are briefly reviewed in Appendix~\ref{app:1}. 
In the MSM, there are eleven physically independent parameters, but only seven parameters of them are observables in low energy experiments.
We summarize 11 independent parameters in different parametrizations in Tab.~\ref{tab:1}. We stress that the choices of independent parameters in some of these parametrizations are not unique, so the parameters listed in Tab.~\ref{tab:1} are just an option. 

In Tab.~\ref{tab:1}, the first five parametrizations contain the seven low energy parameters $m_2$, $m_3(m_1)$, $\theta_{12}$, $\theta_{13}$, $\theta_{23}$, $\delta_{\rm CP}$, $\sigma$, where $m_2 \equiv \sqrt{\Delta m^2_{21}}$ and $m_3 \equiv \sqrt{\Delta m^2_{31}}$ in the NH scenario ($m_2 \equiv \sqrt{-\Delta m^2_{32}}$ and $m_1 \equiv \sqrt{-\Delta m^2_{32} - \Delta m^2_{21}}$ in the IH case). 
The five parameters $\Delta m^2_{21}$, $\Delta m^2_{31}\ (\Delta m^2_{32})$, $\theta_{12}$, $\theta_{13}$, $ \theta_{23}$ have been well-measured in neutrino oscillation experiments \cite{Esteban:2024eli}, but the Dirac phase $\delta_{\rm CP}$ has not been precisely measured and the Majorana phase $\sigma$ is unknown so far, and the rest four parameters from the heavy sector may be constrained in the future high energy experiments. 
The last two parametrizations directly contain only two low energy parameters $m_2$, $m_3(m_1)$, and the rest $\theta_{12}$, $\theta_{13}$, $\theta_{23}$, $\delta_{\rm CP}$, $\sigma$ are not directly presented but constitute new parameters with four high energy parameters.

All parametrizations, except EMSM, Blennow--Fernandez-Martinez and Bi-Unitary parametrizations have used the approximate seesaw formula $M_\nu \approx - M_{\rm D}^{} M_{\rm R}^{-1} M_{\rm D}^{T}$, so their applications might be limited if the high-precision analysis is required in the future study.
The Blennow--Fernandez-Martinez and Bi-Unitary parametrizations are exact and analytical, but their physical meanings are not clear enough.
However, the EMSM contains the advantages of other parametrizations, which is exact, analytical, physical and clear as shown in section~\ref{sec:EMSM}.

\begin{table}[t!]
	\centering
	\renewcommand{\arraystretch}{1.2}
	\begin{tabular}{ll}
		\hline\hline
		parametrization & 11 independent parameters in parametrizations of MSM \\
		\hline
		EMSM & $m_2$, $m_3(m_1)$, $\theta_{12}$, $\theta_{13}$, $\theta_{23}$, $\delta_{\rm CP}$, $\sigma$, $M_1$, $M_2$, $\theta_{14}$, $\delta_{14}$
		\\\hline
		Casas--Ibarra & $m_2$, $m_3(m_1)$, $\theta_{12}$, $\theta_{13}$, $\theta_{23}$, $\delta_{\rm CP}$, $\sigma$, $M_1$, $M_2$, $|\omega|$, $\arg\omega$ 
		\\\hline
		Natural Reconstruction & $m_2$, $m_3(m_1)$, $\theta_{12}$, $\theta_{13}$, $\theta_{23}$, $\delta_{\rm CP}$, $\sigma$, $M_1$, $M_2$, $|a_1|$, $\arg{a_1}$ 
		\\\hline
		Blennow--Fernandez-Martinez & $m_2$, $m_3(m_1)$, $\theta_{12}$, $\theta_{13}$, $\theta_{23}$, $\delta_{\rm CP}$, $\sigma$, $M_1$, $M_2$, $\theta_y$, $\theta_z(\theta_x)$,
		\\\hline
		Modified Casas--Ibarra--Ross & $m_2$, $m_3(m_1)$, $\theta_{12}$, $\theta_{13}$, $\theta_{23}$, $\delta_{\rm CP}$, $\sigma$,  $|P_{13}|$, $|P_{23}|$, $|P_{12}|$, $\arg{P_{12}}$ 
		\\\hline
		Bi-Unitary & $m_2$, $m_3(m_1)$, $M_1$, $M_2$, $\theta_{1}$, $\theta_{2}$, $\theta_{3}$, $\delta_{L}$, $\gamma_{L}$, $\theta_R$,  $\gamma_{R}$ 
		\\\hline
		\multirow{2}{*}{Vector Representation} & $m_2$, $m_3(m_1)$, $M_1$, $M_2$, $u_{e1}$, $u_{\mu 1}$, $|u_{e 2}|$, $ |u_{\mu 2}|$, \\ & $\arg u_{e 2}$, $\arg u_{\mu 2}$, $\arg u_{\tau 2}$
		\\\hline\hline
	\end{tabular}
	\caption{A summary of the eleven independent parameters in the EMSM and other parametrizations of the MSM. $m_3(m_1)$ denotes the NH (or IH) scenario. We inherit conventions in \cite{Guo:2006qa} for names of some of these parametrizations. Note that the set of independent parameters selected in each parametrization is not unique, and that listed above is just one option.}
	\label{tab:1}
\end{table}

\section{Phenomenological implications} \label{sec:pheimp}

We further employ the EMSM parametrization in a few phenomenological implications.
We will apply this parametrization in the studies of directly and indirectly experimental searches for RH neutrinos including charged lepton flavor violations (cLFVs), non-unitarity effects of the lepton mixing and other experiments in section \ref{sec:mu_e_sector}.
The parameter space in these studies will be presented. 
The implication in leptogenesis will be studied in section \ref{sec:leptogenesis}.

\subsection{Experimental constraints} \label{sec:mu_e_sector}

We firstly study the current constrains and future sensitivities on the parameter space of EMSM through cLFV experiments and other directly or indirectly experimental searches for RH neutrinos.
Subsequently, we will analyze the constraints on non-unitarity effects induced by MSM from the updated data of the neutrino oscillations and electroweak precision test (EWPT).

In the cLFV studies, the $\mu$-$e$ sector provides strong constraints in the current and foreseeable future experiments \cite{Antusch:2006vwa,Blennow:2023mqx}.
We mainly focus on the $\mu$-$e$ sector in the EMSM through cLFV experiments.
In the MSM framework, cLFVs are triggered by two RH neutrinos via the active-sterile mixing.  
Hence, we will study these rare processes about the $\mu$-$e$ sector in the EMSM, mainly including radiative decay, $\mu\rightarrow e \gamma$, 3 body decay, $\mu\rightarrow eee$, $\mu$ to $e$ conversion in nucleus ${\rm N}$, $\mu+ {\rm N} \rightarrow e+ {\rm N}$.
The first rare process is the charged lepton radiative decay, and its branching ratio is defined as $Br(\mu\to e\gamma)\equiv\Gamma(\mu\to e\gamma)/\Gamma_\mu$, where $\Gamma_\mu$ is the total muon decay width.
The branching ratio induced by two heavy neutrinos is presented as \cite{Minkowski:1977sc,Marciano:1977wx,Cheng:1980tp,Lim:1981kv,Langacker:1988up},
\begin{align}
	Br(\mu\rightarrow e\gamma)&=\frac{3\alpha_{\rm em}}{2\pi}\left|\sum_{i=1}^3U_{\mu i}^*U_{ei}\times G_{\gamma}\left(\frac{m_i^2}{M_{W}^2}\right)+\sum_{j=1}^2 R_{\mu j}^*R_{ej}\times G_{\gamma}\left(\frac{M_j^2}{M_{W}^2}\right)\right|^2 \, ,
\end{align}
where $G_\gamma(x)$ is a 1-loop function about photons as Eq.~\eqref{eq:Ggamma} shown in Appendix~\ref{app:loopfunc}.
The branching ratio of the lepton 3-body decay $\mu \to eee$ is given by \cite{Ilakovac:1994kj},
	\begin{align}  
		\begin{aligned}
			Br(\mu\to eee) =  \frac{\alpha_{\rm w}^4}{24576 \pi^3} \frac{m_\mu^4}{M_W^4} \frac{m_\mu}{\Gamma_\mu} \times &\left\{  2 \left| (F_{\frac{1}{2}{\rm Box}+Z}^{\mu eee}-F_{Z-\gamma}^{\mu e}) \right|^2 + \left| F_{Z-\gamma}^{\mu e} \right|^2    + 32s_{\rm w}^4 r^{\mu e} \left| G_\gamma^{\mu e} \right|^2       \right.\\
			& \left.\ \ +  16s_{\rm w}^2 \text{Re} \left[ G_\gamma^{\mu e*} (F_{\frac{1}{2}{\rm Box}+Z}^{\mu eee}-\frac{3}{2}F_{Z-\gamma}^{\mu e}) \right]  \right\} \,,  	
		\end{aligned}
	\end{align}
where $\mu, e$ are muon and electron flavor indices respectively, and $r^{\mu e}=\ln(m_\mu^2/m_e^2)-11/4$. Here, $F$ and $G$ are some functions about formal factor depending on $\mathcal{U}_{\alpha i}$, $\mathcal{U}_{\beta i}$, $R_{\alpha j}$, $R_{\beta j}$, $m_\alpha$, $m_\beta$, $M_1$, $M_2$ as Eq.~\eqref{eq:Ggammaalphabeta} and Eq.~\eqref{eq:F} in Appendix~\ref{app:loopfunc}.
Finally, denoting the atomic number of nucleus N by $Z$, the $\mu$-$e$ conversion rate in light nucleus ($\alpha_{\rm em} Z\ll 1$) is given in \cite{Alonso:2012ji},
\begin{align}
	R_{\mu \to e} = \frac{G_F^2 \alpha_{\rm w}^2 \alpha_{\rm em}^3 m_\mu^5 Z_{\text{eff}}^4}{8 \pi^4 \Gamma_{\text{capt}} Z} F_p^2 \Bigg| \sum_{j=1}^2 R_{ej}^{}R_{\mu j}^* \Big[ Z ( 2 F_u + F_d ) + (A - Z) ( F_u + 2 F_d ) \Big] \Bigg|^2  \,, \label{eq:Rmue}
\end{align}
where $F_u$ and $F_d$ are functions depending on $M_j^2/M_W^2$ as Eq.~\eqref{eq:Fud} in Appendix~\ref{app:loopfunc}.
We take nucleus $_{22}^{48}$Ti to determine the form factor $|F_p|=0.54$, effective atomic number $Z_{\rm eff}=17.6$ and the caption rate $\Gamma_{\rm capt}=2.59\times 10^6\ {\rm s}^{-1}$  \cite{Kitano:2002mt,Suzuki:1987jf}.
The current and further upper limits for the branching ratios of three cLFV processes via relavant experiments are listed in Tab.~\ref{tab:mu_e_sector}.
\begin{table}[t]  
	\centering  
	\begin{spacing}{1.5} 
		\begin{tabular}{ccc}  
			\hline\hline  
			~\quad Processes \quad~ & ~\quad Current bounds \quad~ & ~\quad Future sensitivities \quad~ \\
			\hline  
			$Br(\mu \to e\gamma)$ & $ 3.1 \times 10^{-13}$\cite{MEGII:2023ltw} & $ 6.0 \times 10^{-14}$\cite{MEGII:2018kmf} \\
			\hline  
			$Br(\mu \to eee)$ & $ 1.0 \times 10^{-12}$ \cite{SINDRUM:1987nra} & $ 1.0 \times 10^{-16}$\cite{Mu3e:2020gyw} \\
			\hline  
			$R_{\mu \to e}$ (Ti) & $ 4.3 \times 10^{-12}$\cite{SINDRUMII:1993gxf} & $ 1.0 \times 10^{-18}$\cite{Hungerford:2009zz, COMET:2009qeh} \\
			\hline\hline  
		\end{tabular}
	\end{spacing}
	\caption{The current upper bounds and future sensitivities for branching ratios of cLFV processes in the $\mu$-$e$ sector.}  
	\label{tab:mu_e_sector}  
\end{table} 
The $\mu$-$e$ conversion might be highly suppressed due to the cancellation inside the square brackets in Eq.~\eqref{eq:Rmue}, which is induced by the opposite signs of $F_u$ and $F_d$. The mixing angle $\theta_{14}$ is hardly constrained around the cancellation point.
In general, this point depends on the mass of heavy neutrinos $M_1,M_2$ and the active-sterile mixing (i.e. $\theta_{14}$).
After simplifying $F_u$ and $F_d$ as Eq.~\eqref{eq:Fudsim} in Appendix~\ref{app:loopfunc} for $M_j^2/M_W^2\gg 1$, we obatin the functional relation of $M_1$, $M_2$ and $\theta_{14}$ at the resonant point,
\begin{align}
	&\left\{R_{e1}^{}R_{\mu 1}^*\ln\left(\frac{M_1^2}{M_W^2}\right)+R_{e2}^{}R_{\mu 2}^*\ln{\left(\frac{M_2^2}{M_W^2}\right)}\right\}\bigg|_{\rm res}
	= \left(R_{e1}^{}R_{\mu 1}^*+R_{e2}^{}R_{\mu 2}^*\right) f(A,Z) \,,
\end{align}
where $f(A,Z)$ is a function determined by atomic nucleus,
\begin{align}
	f(A,Z):=\frac{ \frac{9}{8} (A - Z) + \left( \frac{9}{8} + \frac{31}{12} s_W^2 \right) Z } { \frac{3}{8} (A - Z) - \left(\frac{3}{8}- \frac{4}{3} s_W^2  \right) Z } \,,
\end{align}
and $f(A,Z)\simeq 8.1$ for $_{22}^{48}{\rm Ti}$.
In the case of degenerate heavy neutrinos, $M_1\simeq M_2$, the resonant point is independent of the mixing matrix $R$ and then the resonant mass is
\begin{align}
	M^2|_{\rm res} = M_W^2 \cdot{\rm Exp} \left( f(A,Z) \right) \,.
\end{align}
Therefore, the resonant point for titanium satisfies $M|_{\rm res}\simeq 3.375\times 10^3$~GeV in the degenerate case of heavy neutrinos .

\begin{figure}[h!]
	\centering
	\subfloat[Current constraints (left) and future sensitivities (right, current constraints shown in light gray) on the heavy neutrino mass $M_1$ and the element of the active-sterile mixing matrix $|U_{eN}|$. $M_1\simeq M_2$ is assumed. The dark gray region below the line $\theta_{14}=0$ refers to the failure of the minimal seesaw model.]
	{\label{fig:constrainsU}
	\begin{minipage}{0.95\linewidth}
		\centering
		\includegraphics[width=\linewidth]{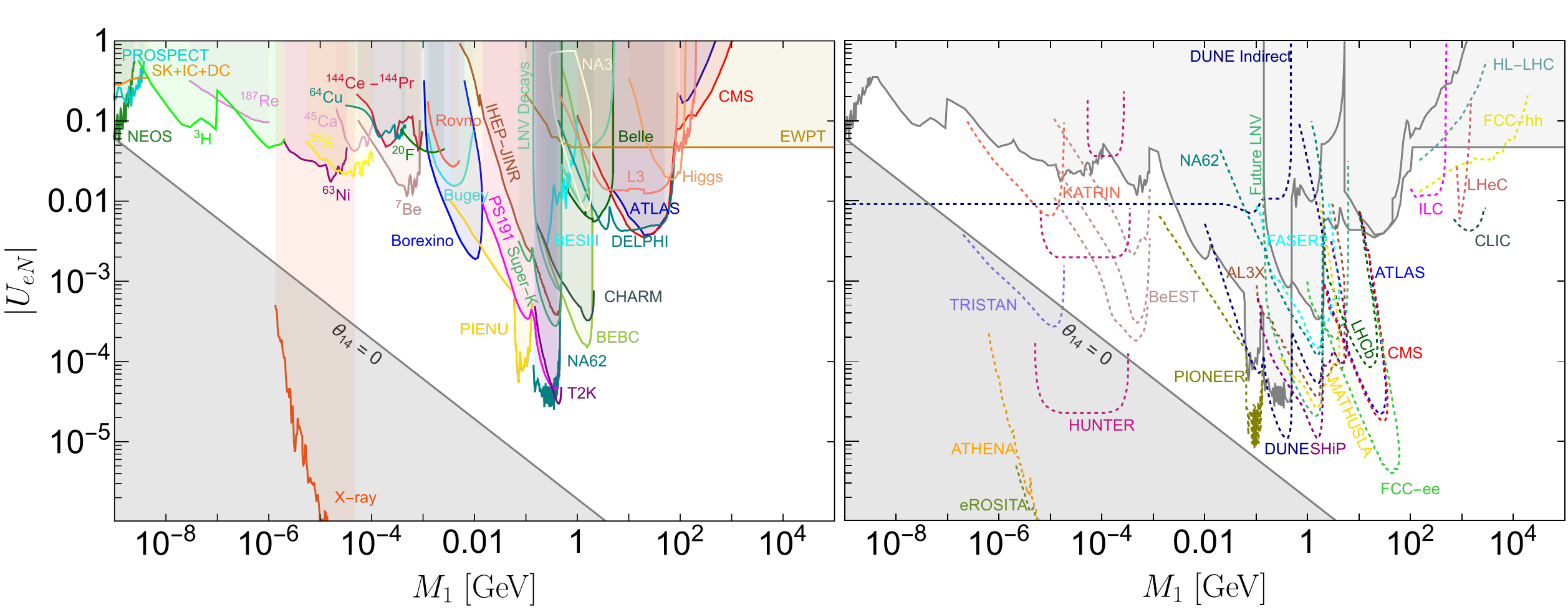}
	\end{minipage}
	} \\
	\subfloat[Current constraints (left) and future sensitivities (right) on the heavy neutrino mass $M_1$ and the active-sterile mixing angle $\theta_{14}$. $M_1\simeq M_2$ and $\sigma=\delta_{14}=0$ are assumed.]
	{\label{fig:constrainstheta}
	\begin{minipage}{0.95\linewidth}
		\centering
		\includegraphics[width=\linewidth]{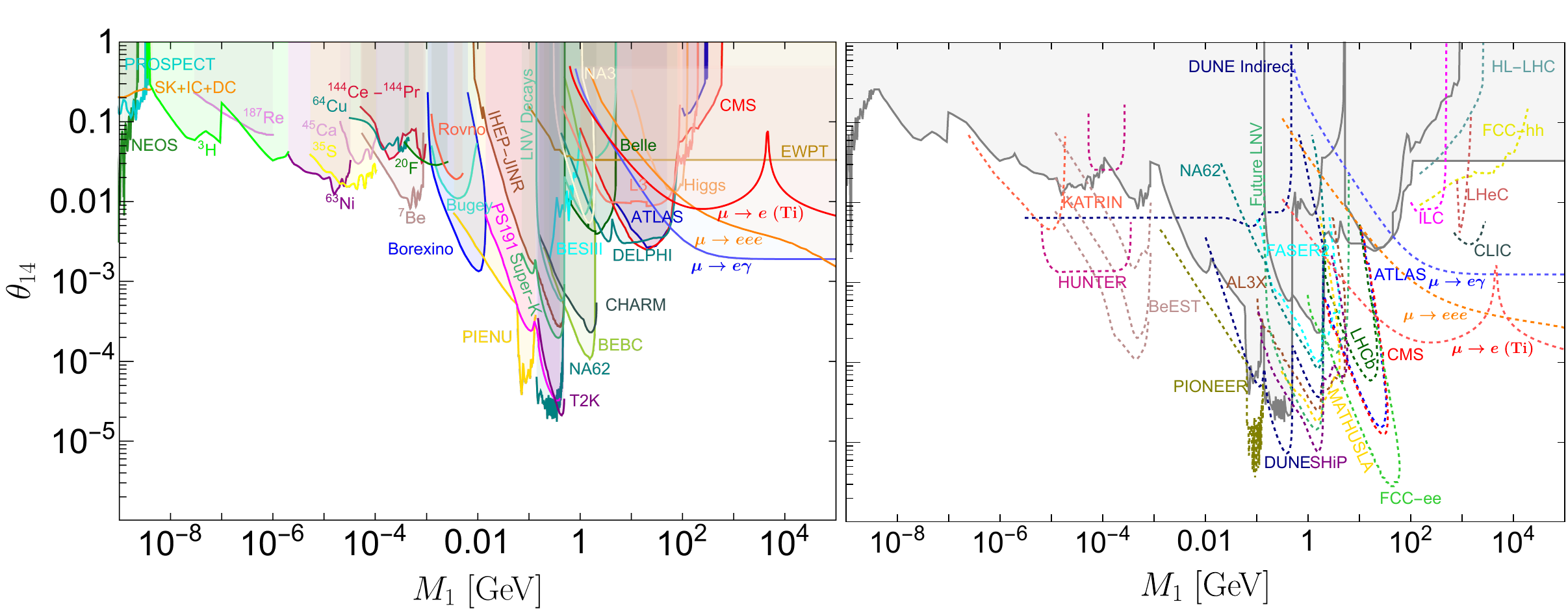}
	\end{minipage}
	}\\
	\subfloat[Relative difference between the exact and approximate active-sterile mixing angle about the heavy neutrino mass $M_1$, where $\delta\theta_{14}/\theta_{14}=|\theta_{14}^{\rm CI}/\theta_{14}^{\rm EMSM} - 1|$. $M_1\simeq M_2$ and $\sigma=\delta_{14}=0$ are assumed.]
	{\label{fig:constrainsdelta}
	\begin{minipage}{0.95\linewidth}
		\centering
		\includegraphics[width=\linewidth]{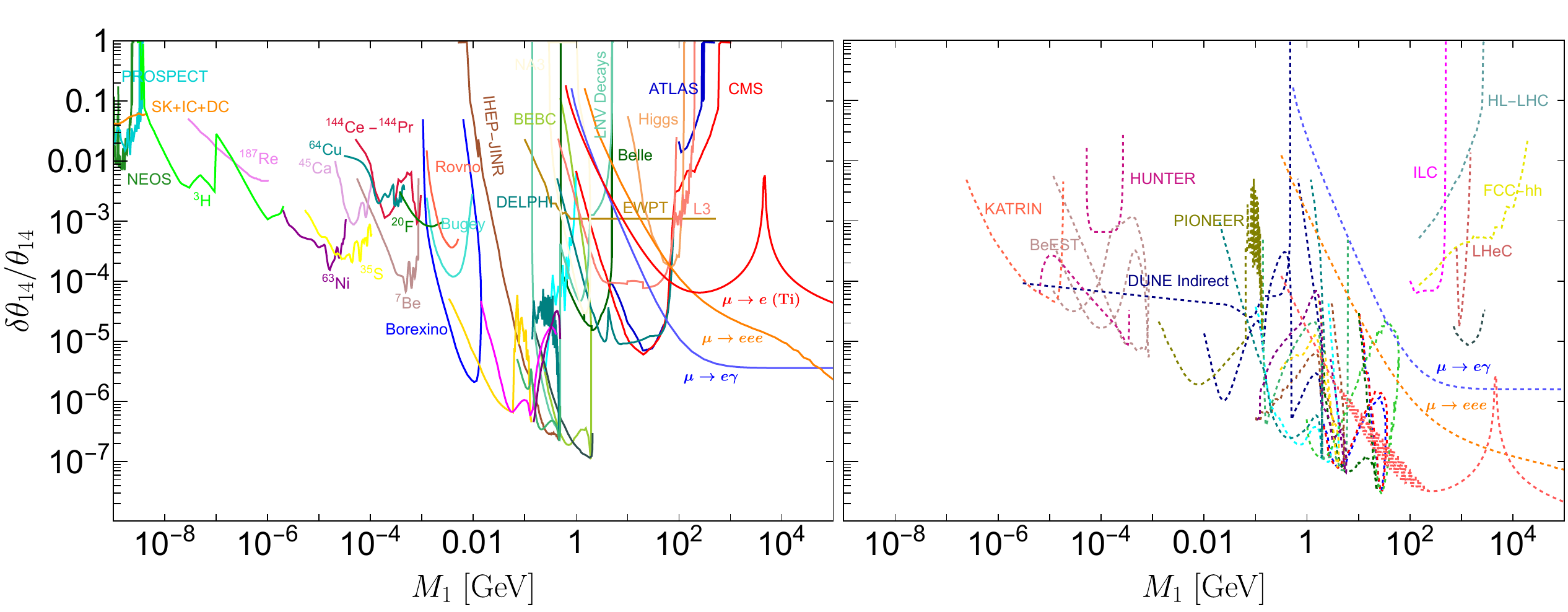}
	\end{minipage}
	}
	\caption{Implication of EMSM in experimental searches and its advantage compared to parametrizations from approximate seesaw formula.}
\end{figure}

The other experimental searches for RH neutrinos include direct and indirect collider searches, meson decays and beam-dump experiments, beta decays and nuclear processes, active-sterile neutrino oscillations, EWPT and astrophysical observations. The data of these experiments is summarized in \cite{Bolton:2019pcu}.
These experimental searches can directly constrain the active-sterile mixing matrix $U$. Assuming that the two RH neutrino masses are quasi-degenerate $M_1\simeq M_2$, the current constrains and future sensitivities on $U_{eN}$ and $M_1$ from these experiments is shown in Fig.~\ref{fig:constrainsU}. 
Based on EMSM, current constraints and future sensitivities on the parameter space of $M_1$ and the exact active-sterile mixing angle $\theta_{14}$ from cLFVs and other experimental searches are exhibited in Fig.~\ref{fig:constrainstheta}. Here, we assume the CP phases $\sigma=\delta_{14}=0$, which reduces degrees of freedom of parameters.
We give Eq.~\eqref{eq:equivalence} to calculate the approximate active-sterile mixing angle $\theta_{14}^{\rm CI}$ in CI parametrization. Actually, active-sterile mixing angles in former different parametrization are the same, because these parametrizations are all derived from Eq.~\eqref{eq:approx_ss}, while our parametrization is based on the exact formula Eq.~\eqref{eq:exact_ss}.
Meanwhile, in Fig.~\ref{fig:constrainstheta}, the pole on the red curve refers to the cancellation of the $\mu$-$e$ conversion in titanium in Eq.~\eqref{eq:Rmue}.
From Fig.~\ref{fig:constrainstheta}, current experiments suggest the active-sterile mixing angle $\theta_{14} \lesssim {\cal O}(10^{-3})$ and future experiments can test $\theta_{14} \sim {\cal O}(10^{-4})$ when the RH neutrino mass is larger than 1~GeV.

To highlight parameter regions where other commonly used parameterizations fail, we use $\delta\theta_{14}/\theta_{14}=|\theta_{14}^{\rm CI}/\theta_{14}^{\rm EMSM} - 1|$ to quantify the relative error of the approximate seesaw formula compared with the exact one, and the result is shown in Fig.~\ref{fig:constrainsdelta}. This error can reach as large as ${\cal O}(0.01\text{-}0.1)$ in some experimental constraints, including the established measurements: NEOS, PROSPECT, SK+IC+DC, $\rm^3H$, $\rm^{187}Re$, $\rm^{144}Ce$ $-\rm^{144}Pr$, Borexino, IHEP--JINR, BEBC, Belle, NA3, LNV Decays, Higgs, L3, ATLAS, CMS, $\mu\to e({\rm Ti})$, $\mu\to eee$, $\mu\to e\gamma$, and future proposal: HUNTER, DUNE Indirect, ILC, HL--LHC, LHeC, FCC--hh, $\mu\to eee$, $\mu\to e\gamma$.
Therefore, we suggest to use EMSM in these experimental constraints/sensitivities.
We further show in Tab.~\ref{tab:constrainstheta} the comparison of constraints on the upper bound of $\theta_{14}$ obtained from our parameterization and other parameterization for the sterile neutrino masses at 1 eV.

\begin{table}[h!]  
	\centering  
	\renewcommand{\arraystretch}{1.2}
	\begin{tabular}{w{c}{3cm}w{c}{2cm}w{c}{2.5cm}w{c}{2.5cm}w{c}{1.5cm}}
		\hline\hline 
		$\theta_{14}$  & NEOS & PROSPECT & SK+IC+DC \\ 
		\hline 
		EMSM & 0.0209 & 0.124 & 0.203 \\
		\hline
		\makecell[c]{Former\\ parametrizations} &  0.0204 & 0.122 & 0.195 \\ 
		\hline
		$\delta\theta_{14}/\theta_{14}$ & 2.47\% & 1.93\% & 4.23\% \\
		\hline\hline  
	\end{tabular}  
\caption{ Constraints on the upper bound of the sterile-active mixing angle $\theta_{14}$ at $M_1= 1$~eV in EMSM and other parameterizations. All former parametrizations, due to the use of approximate seesaw formula,  induce an error of $\mathcal{O}(0.01)$--$\mathcal{O}(0.1)$.}
	\label{tab:constrainstheta}  
\end{table}


We further check the constraints on active-sterile mixing angles in EMSM from constraints of non-unitarity effects from neutrino oscillations and electroweak precision test (EWPT).
The $3\times 3$ mixing matrix is not exactly an unitary matrix in the presence of RH neutrinos. 
In the MSM, the heavy neutrinos $N_1, N_2$ participate in the flavor mixing through nonzero active-sterile mixing angles, which induces the non-unitarity effects \cite{Xing:2007zj,Xing:2011ur,Escrihuela:2015wra,Li:2015oal}. 
As discussed in section~\ref{sec:EMSM}, the non-unitarity of the effective flavor mixing matrix $U=AU_0$ in Eq.~\eqref{eq:mathU} is due to the deviation of $A$ from unit matrix $I$ in Eq~\eqref{eq:A}.
Therefore, we will use $I-A$ to evaluate the non-unitary effects of the lepton mixing.
The non-unitarity behavior of neutrinos can be constrained by global fits from the neutrino oscillation experiments \cite{Forero:2021azc} and electroweak precision test (EWPT) experiments \cite{Blennow:2023mqx} as shown in Tab.~\ref{tab:non-unitarity1}, where $\alpha_{\alpha \beta}\equiv A_{\alpha\beta}$ in Eq.~\eqref{eq:A} \footnote{Actually, in some literature, the definition of $\alpha$ is unitary transformation for $A$, so $\alpha^\dagger\alpha=A^\dagger A$ or $\alpha\alpha^\dagger=AA^\dagger$ is stricter.}
and the data within 95\% confidence level (C.L.) from the neutrino oscillation is obtained through paper \cite{Forero:2021azc}.
Here, it is clear that the constraints from the EWPT are stricter than the neutrino oscillation.
The constraints on $\alpha_{\alpha \beta}$ are straightforwardly converted into the constraints on active-sterile mixing angles $\theta_{ik}$ and the results are shown in Tab.~\ref{tab:non-unitarity2}, where we consider two scenarios for RH neutrino masses, the degeneracy ($M_1\simeq M_2$) and the hierarchy ($M_1\ll M_2$).
Tab.~\ref{tab:non-unitarity2} shows that the active-sterile mixing angles are restricted mostly to lower than $10\degree$ and $1\degree$ through the neutrino oscillation and the EWPT, respectively, though the complex phases are still unconstrained.

\begin{table}[h!]  
	\centering  
	\renewcommand{\arraystretch}{1.2}
	\begin{tabular}{cccc}  
		\hline\hline  
		Non-unitarity para. &  ~\qquad$\nu$ oscillation \qquad~ & ~\qquad EWPT \qquad~ \\
		\hline  
		$1-\alpha_{ee}$ & $3.9 \times 10^{-2}$ & $9.4 \times 10^{-6}$ \\
		$1-\alpha_{\mu \mu}$ & $6.2 \times 10^{-3}$ & $1.3 \times 10^{-4}$ \\
		$1-\alpha_{\tau \tau}$ & $1.4 \times 10^{-1}$ & $2.1 \times 10^{-4}$ \\
		\hline
		$|\alpha_{\mu e}|$ & $1.8 \times 10^{-2}$ &  $2.4 \times 10^{-5}$ \\
		$|\alpha_{\tau e}|$ & $4.2 \times 10^{-2}$ & $4.4 \times 10^{-5}$ \\
		$|\alpha_{\tau \mu}|$ & $1.1 \times 10^{-2}$ & $2.6 \times 10^{-4}$ \\
		\hline\hline  
	\end{tabular}  
	\caption{ Upper bounds on the non-unitarity effects of neutrinos by global fits from neutrino oscillation experiments \cite{Forero:2021azc} and EWPT experiments \cite{Blennow:2023mqx} at 95\%C.L., where $\alpha\equiv A$ in Eq.~\eqref{eq:A}.}
	\label{tab:non-unitarity1}  
\end{table}

\begin{table}[h!]  
	\centering 
	\begin{spacing}{1.2}
	\begin{tabular}{cccccc}  
		\hline\hline
		Mixing  & \multicolumn{2}{c}{$\nu$ oscillation } && \multicolumn{2}{c}{EWPT } \\
		\cline{2-3} \cline{5-6}
		angles & Degeneracy & Hierarchy && Degeneracy & Hierarchy \\
		\hline
		$\theta_{14}$ & \multirow{2}{*}{$4.4\times 10^{-2}\, (2.5\degree)$} & $9.0\times 10^{-2}\, (5.2\degree)$ && \multirow{2}{*}{$3.0\times 10^{-3}\, (0.17\degree)$} & $4.3\times 10^{-3}\, (0.25\degree)$\\
		\cline{3-3}\cline{6-6}
		$\theta_{15}$ &  & --- &  && --- \\
		\hline
		$\theta_{24}$ & \multirow{2}{*}{$7.9\times 10^{-2}\, (4.5\degree)$} & $1.1\times 10^{-1}\, (6.4\degree)$ && \multirow{2}{*}{$1.1\times 10^{-2}\, (0.64\degree)$} & $1.6\times 10^{-2}\, (0.92\degree)$\\
		\cline{3-3}\cline{6-6}
		$\theta_{25}$ &  & --- &  && --- \\
		\hline
		$\theta_{34}$ & \multirow{2}{*}{$1.0\times 10^{-1}\, (5.9\degree)$} & $1.8\times 10^{-1}\, (10.4\degree)$ && \multirow{2}{*}{$7.2\times 10^{-3}\, (0.41\degree)$} & $8.9\times 10^{-3}\, (0.51\degree)$\\
		\cline{3-3}\cline{6-6}
		$\theta_{35}$ &  & --- &  && --- \\
		\hline\hline
	\end{tabular}  		
    \end{spacing}
	\caption{ Upper bounds on the active-sterile mixing angles converted from Table~\ref{tab:non-unitarity1}.  Two scenarios for RH neutrino masses, the degeneracy ($M_1 \simeq M_2$) and the hierarchy ($M_1 \ll M_2$), are considered. }  
	\label{tab:non-unitarity2}
\end{table}    

\subsection{Leptogenesis}\label{sec:leptogenesis}
The observations about the cosmic microwave background (CMB) \cite{Planck:2018vyg} and the Big-Bang nucleosynthesis (BBN) \cite{ParticleDataGroup:2024cfk} reveal that baryons (or matter) are more than antibaryons (or antimatter), i.e. the matter-antimatter asymmetry.
Baryogenesis via leptogenesis \cite{Fukugita:1986hr} is one of the most important mechanisms to explain the matter-antimatter asymmetry.
In this section, we will study the parameter space of the EMSM parametrization that can realize successful leptogenesis at low energy scale $10^3$~GeV -- $10^8$~GeV.

The Boltzmann equations of heavy neutrinos $N_{N_i} (i=1,2)$ and $B-L$ asymmetry $N_{B-L}$ with flavor effects \cite{Barbieri:1999ma} taken into account are given by
\begin{align}
	\frac{{\rm d}N_{N_i}}{{\rm d}z} &= -D_i \left(N_{N_i}^{} - N_{N_i}^{\mathrm{eq}}\right)\,, \label{eq:Boltz1} \\  
	\frac{{\rm d}(N_{B-L})_{\alpha \alpha}}{{\rm d}z} &= \sum_{i}\Bigg[\left(-\varepsilon_{i \alpha}D_i\, (N_{N_i}^{} - N_{N_i}^{\mathrm{eq}}) \right) - K_{i\alpha}W_i \sum_{\beta=e,\mu,\tau} \left(C_{\alpha\beta}^l+C_{\beta}^H\right)(N_{B-L})_{\beta\beta}\Bigg] \,, \label{eq:Boltz2}
\end{align}
where $z\equiv M_1/T$, $\alpha$ is a flavor index of charged leptons, $D_i$ and $W_i$ are $N_i$'s decay and washout terms, respectively, shown in \cite{Buchmuller:2004nz}, $K_{i\alpha}$ is the projection probabilities between the mass and flavor states \cite{Nardi:2006fx}, and, $C^l_{\alpha\beta}$ and $C_\beta^H$ are spectator effects factors shown in \cite{Abada:2006ea,Nardi:2006fx}.
The number densities $N_{N_i}$ and $N_{B-L}$ are normalized to the comoving volume containing one photon.  
Here, $\varepsilon_{i\alpha}$ represents the CP asymmetry defined as
\begin{align}
	\begin{aligned}
		\varepsilon_{i\alpha}&\equiv\frac{\Gamma(N_i^{}\rightarrow \ell_{\alpha}^{}+H)-\Gamma(N_i^{}\rightarrow\Bar{\ell}_{\alpha}+\Bar{H})}{\sum_{\alpha}^{}\left[\Gamma(N_i^{}\rightarrow\ell_{\alpha}+H)+\Gamma(N_i^{}\rightarrow\Bar{\ell}_{\alpha}^{}+\Bar{H}^{})\right]} \,,
	\end{aligned} 
\end{align}
which is generated by the interference between the tree and 1-loop effects. We denotes $\varepsilon_\alpha=\sum_i \varepsilon_{i\alpha}$.
In the highly degenerate mass regions where the RH neutrino mass difference is comparable with their decay widths, the CP asymmetry is resonantly increased via enhanced heavy-neutrino self-energy effects in resonant leptogenesis,
\begin{align}
	\begin{aligned}
			\varepsilon_{i \alpha} =\sum_{j\neq i}  \frac{\text{Im} \left[ (Y_{\nu}^*)_{\alpha i} (Y_{\nu})_{\alpha j}^{} (Y_{\nu}^\dagger Y_{\nu})_{ij}^{} \right] + \frac{M_i}{M_j} \text{Im} \left[ (Y_{\nu}^*)_{\alpha i} (Y_{\nu})_{\alpha j}^{} (Y_{\nu}^\dagger Y_{\nu})_{ij}^{*} \right]}{(Y_{\nu}^\dagger Y_{\nu})_{ii}^{} \, (Y_{\nu}^\dagger Y_{\nu})_{jj}^{}}\cdot \frac{\Delta M_{ij}^2\, M_i^{} \Gamma_j^{}}{(\Delta M_{ij}^2)^2 + M_i^2 \Gamma_j^2} \,,
	\end{aligned} \label{eq:epsilonre} 
\end{align}
where $\Delta M_{ij}^2=M_i^2-M_j^2$, $\Gamma_i=\frac{(Y_\nu^\dagger Y_\nu)_{ii}}{8\pi}M_i$ is the $N_i$  decay width at tree level, and the second term is a resonant-enhancement factor.
The matter-antimatter asymmetry is characterized by the ratio of the baryon-antibaryon number density difference to the photon number density
\begin{align}
	\eta_{B}\equiv\frac{n_B-n_{\bar{B}}}{n_\gamma}=\frac{c_S}{f} N_{B-L} \,,
\end{align}
where $c_S$ is the fraction of $B-L$ asymmetry converted to $B$ asymmetry by sphaleron process shown in \cite{Khlebnikov:1988sr,Harvey:1990qw}, $f$ is a dilution factor and $f=n_{\rm rec}^\gamma a_{\rm rec}^3/(n_*^\gamma a_*^3)=g_\star/g_{\rm rec}$.
In the MSM and standard thermal-universe history, $c_S=\frac{28}{79}$ and $f\simeq 27.65$.
The matter-antimatter asymmetry $\eta_B$ was measured most precisely in the CMB observation satisfying $\eta_B=(6.12\pm 0.04)\times 10^{-10}$ in $68\%\ {\rm C.L.}$ \cite{Planck:2018vyg}.

We apply the EMSM parametrization in leptogenesis as below. 
As discussed in section~\ref{sec:EMSM}, $Y_\nu=\sqrt{2} M_{\rm D}/v$ and the exact expression of $M_{\rm D}$ is Eq.~\eqref{eq:MD}.
After inputting low energy parameters and taking heavy neutrinos masses $M_1,M_2$ and fundamental parameters $\theta_{14},\delta_{14}$ of EMSM, we can calculate the unitary matrix $\mathcal{U}$ of Eq.~\eqref{eq:mathU} through the method in section~\ref{sec:EMSM}, and then obtain $Y_\nu$ from Eq.~\eqref{eq:MD}.
Finally, the CP asymmetry is calculated via Eq.~\eqref{eq:epsilonre}, and subsequently the baryon-antibaryon asymmetry is evaluated by Eq.~\eqref{eq:Boltz1} and Eq.~\eqref{eq:Boltz2}. 

Numerically, the \textit{python} package ULYSSES \cite{Granelli:2020pim,Granelli:2023vcm} can successfully calculate the evolution of particle number densities and give the final $\eta_B$. The procedure is arranged in the following. 
Firstly, we should obtain the Yukawa coupling $Y_{\nu}$. 
Hence, taking the $\nu_1-N_1$  mixing angle $\theta_{14}=10^{-6},10^{-8}$, the heavy neutrino mass $M_1\in [10^3,10^8]$~GeV and the degeneracy of heavy neutrinos $(M_2-M_1)/M_1\in [10^{-12},10^{-4}]$, and inputting known low energy parameters, we scan the 2 phase angles  $\delta_{14}, \sigma \in (-\pi,\pi]$ and then choose the values of phase angles to make $\varepsilon_{\rm max} \equiv \max\{-\varepsilon_e,-\varepsilon_\mu,-\varepsilon_\tau\}$ maximized.
Next, we use the ULYSSES to calculate the matter-antimatter asymmetry $\eta_B$ after inputting physical parameters and these Yukawa couplings at different heavy neutrino mass $M_1$ and degeneracy $(M_2-M_1)/M_1$.
We use the mode 2RESsp (considering spectator effects) in resonant leptogenesis.\footnote{Here, we have used the python GNU parallel from the website: https://doi.org/10.5281/zenodo.15071920.}

\begin{figure}[h] 
	\centering 
	\includegraphics[width=.9\textwidth]{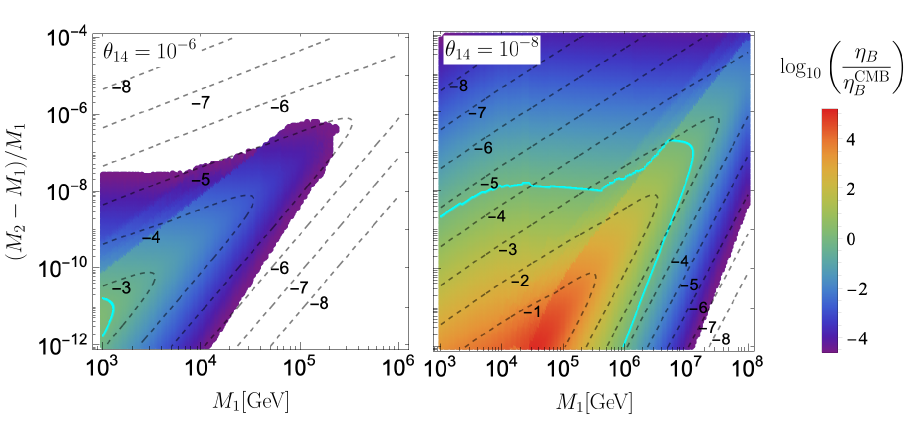} 
	\caption{The heat map of the matter-antimatter asymmetry $\eta_B$ and the counter plot of the CP asymmetry $\varepsilon_{\rm max}$ along the heavy neutrino mass $M_1$ and the degeneracy parameter $(M_2-M_1)/M_1$ when the active-sterile mixing angle $\theta_{14}=10^{-6}$ (left) and $10^{-8}$ (right). Here, $\eta_B^{\rm CMB}=6.12\times 10^{-10}$ is the best-fit value of the CMB observation.
	The dashed curves are counter lines of $\log_{10}\varepsilon_{\rm max}$ and corresponding values are marked along the curves.
	The colors represent the values of $\log_{10}\eta_B/\eta_B^{\rm CMB}$, where the narrow cyan band satisfies the CMB observation $\eta_B\in (6.12\pm 0.20)\times 10^{-10}$ in $5\sigma$ C.L..}
	\label{fig:asymmetry}  
\end{figure}

\begin{figure}[h] 
	\centering 
	\includegraphics[width=1.\textwidth]{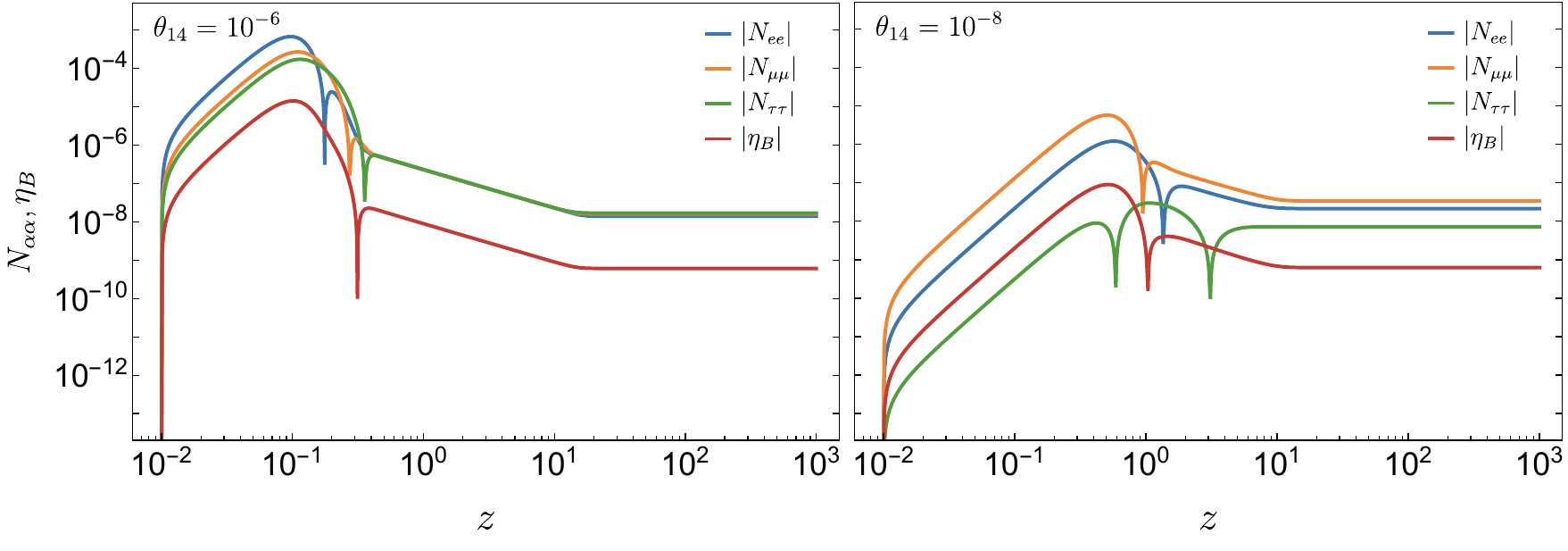} 
	\caption{ The matter evolutions satisfying CMB observations for resonant leptogenesis. Here, $N_{\alpha\alpha}\equiv(N_{B-L})_{\alpha\alpha}$. Left panel: $\theta_{14}=10^{-6}$, $M_1=10^3$~GeV, $|M_2/M_1-1|\simeq 1.74\times 10^{-12}$, $\delta_{14}=1.72\pi$ and $\sigma=1.52\pi$; right panel: $\theta_{14}=10^{-8}$, $M_1=10^3$~GeV, $|M_2/M_1-1|\simeq 2.09\times 10^{-9}$, $\delta_{14}=0.72\pi$ and $\sigma=1.64\pi$.}
	\label{fig:evolution}  
\end{figure}

The matter-antimatter asymmetry $\eta_B$ and the CP asymmetry $\varepsilon_{\rm max}$ are shown in Fig.~\ref{fig:asymmetry}, and the representative figures of matter evolutions are given in Fig.~\ref{fig:evolution}.
As we shown in Fig.~\ref{fig:asymmetry}, leptogenesis at low energy scale can successfully generate the matter-antimatter asymmetry satisfying the CMB observation.

\section{Conclusions} \label{sec:conclusion}

The seesaw mechanism provides a natural way in the explanation of tiny neutrino masses by introducing heavy mediators. In the minimal seesaw model (MSM), only two right-handed (RH) neutrinos are introduced beyond the Standard Model, two active neutrinos gain small masses via the active-sterile mixing, and the lightest active neutrino keeps massless. 
Several parametrizations of the MSM have been provided successively in the literature. Most of them employ the approximate seesaw formula $M_{\nu}\simeq -M_{\rm D}^{}M_{\rm R}^{-1}M_{\rm D}^{T}$ under the assumption $M_{\rm D}\ll M_{\rm R}$.
In this work, we propose a parametrization without considering this approximation, and we call it the Exact parametrization of the MSM (EMSM).

In the MSM, we can parametrize the most general $5\times 5$ (3 active + 2 sterile) neutrino mass matrix in terms of 11 independent parameters.
In the EMSM parametrization, these parameters include 6 neutrino oscillation parameters $\Delta m^2_{21}$, $\Delta m^2_{31}$, $\theta_{12}$, $\theta_{13}$, $\theta_{23}$, $\delta_{\rm CP}$, 1 mass parameter $m_{ee}$ used in $0\nu 2 \beta$ decay processes, 2 sterile neutrino masses $M_1$, $M_2$, 1 active-sterile mixing angle $\theta_{14}$ and 1 relevant phase $\delta_{14}$ for the NH or IH case.
The other parameters, e.g. other active-sterile mixing angles and phases and the Dirac mass matrix, can be represented by these 11 parameters exactly without any approximation. The results are given in the last part of section~\ref{sec:EMSM}. 
For better performing advantages of this parametrization, we compared it with other parametrizations in section~\ref{sec:comparepara}.
In particular, correlations with Casas-Ibarra and Natural Reconstruction parametrizations are presented.

We applied this parametrization in several phenomenological studies in section~\ref{sec:pheimp}.
The active-sterile mixing angle $\theta_{14}$ is constrained to $\theta_{14}\lesssim \mathcal{O}(10^{-3})$ from directly and indirectly experimental searches if the RH neutrino masses are larger than 1~GeV.
In some experimental searches, the relative error of active-sterile mixing in parametrizations from the approximative seesaw formula can reach percent level. Regions of these large errors motivate the use of exact seesaw formula in the presentation of experimental constraints and sensitivities.
Meanwhile, we employed the EMSM to non-unitarity effects and then limited $\nu_i-N_k$ mixing angles $\theta_{ik}\lesssim \mathcal{O}(1\degree)$.
Finally, we showed examples of successful leptogenesis through the EMSM parametrization with sterile neutrino masses below $10^8$~GeV.

The EMSM parametrization is exact and analytical and its parameters have clearly physical meanings.
This parametrization will be helpful to high-precision measurements in the near future.

\section*{Acknowledgment}

ZQC is especially grateful to Z.Z. Xing for the initial guidance on neutrino theories. We also thank G.X. Fang and J. Turner for their useful discussions. 
This work is supported by National Natural Science Foundation of China (NSFC) under Grants Nos. 12205064, 12347103 and Zhejiang Provincial Natural Science Foundation of China under Grant No. LDQ24A050002.

\section*{Appendix}
\appendix

\section{Several parametrizations of the minimal seesaw model}\label{app:1}
\textbf{Casas--Ibarra parametrization \cite{Casas:2001sr,Ibarra:2003up}}\\
The well-known Casas-Ibarra parametrization \cite{Casas:2001sr} is   
\begin{equation}
	M_{\rm D} \simeq iV_0 \sqrt{D_\nu}\ \Omega\, \sqrt{D_N} \,, \label{eq:CIR}
\end{equation}
where the approximate seesaw formula, $M_{\nu} \simeq -M_{\rm D} D_N^{-1} M_{\rm D}^{T}$, has been employed and $\Omega$ is an orthogonal matrix.
Applying this parametrization to the special case of only two RH neutrinos\cite{Ibarra:2003up}, called the Casas-Ibarra-Ross parametrization, $\Omega$ satisfies $\Omega\,\Omega^{T}=\text{diag}\{0,1,1\}$ or $\{1,1,0\}$ in the NH or IH respectively and $\Omega^{T}\Omega=\mathbf{I}_{2\times 2}$, so it can be represented as 
\begin{eqnarray}\label{equ:omega}
\Omega = \begin{pmatrix}
0 & 0 \\
\cos\omega  & -\sin\omega \\
\xi\sin\omega & \xi\cos\omega
\end{pmatrix} \; \text{(NH)}\,, \quad \text{or} \quad \
\Omega=\begin{pmatrix}
\cos\omega  & - \sin\omega \\
\xi\sin\omega & \xi\cos\omega\\
0 & 0 \\
\end{pmatrix} \; \text{(IH)}\,,
\end{eqnarray}
where $\xi=\pm 1$ and $\omega = |\omega| e^{i \arg \omega}$.
The entries of $M_{\rm D}$ for the NH and IH cases are shown, 
\begin{align}
	(M_{\rm D}^{})_{\alpha 1}^{} &\simeq
	\begin{cases}
		+i\sqrt{M_1} \left((V_0)_{\alpha 2} \sqrt{m_2} \cos \omega + \xi\  (V_0)_{\alpha 3} \sqrt{m_3} \sin \omega \right) & \text{(NH)} \\
		+i\sqrt{M_1} \left((V_0)_{\alpha 1} \sqrt{m_1} \cos \omega + \xi\  (V_0)_{\alpha 2} \sqrt{m_2} \sin \omega \right) & \text{(IH)} 
	\end{cases}\,, \\
	(M_{\rm D}^{})_{\alpha 2}^{} &\simeq
	\begin{cases}
		-i\sqrt{M_2} \left((V_0)_{\alpha 2} \sqrt{m_2} \sin \omega -\xi\  (V_0)_{\alpha 3} \sqrt{m_3} \cos \omega \right) & \text{(NH)} \\
		-i\sqrt{M_2} \left((V_0)_{\alpha 1} \sqrt{m_1} \sin \omega -\xi\  (V_0)_{\alpha 2} \sqrt{m_2} \cos \omega \right) & \text{(IH)}
	\end{cases} \,,
\end{align}
where $\alpha = e,\mu,\tau$ is a flavor index.
This parametrization contains 11 real parameters in total, i.e. 7 parameters in light neutrinos (2 masses $m_2$, $m_3(m_1)$, 3 mixing angles $\theta_{12}$, $\theta_{13}$, $\theta_{23}$, 1 Dirac phase $\delta_{\rm CP}$ and 1 Majorana phase $\sigma$), and 4 new independent parameters from heavy neutrinos ($M_1$, $M_2$, 
$|\omega|$ and $\arg{\omega}$), as summarized in Table~\ref{tab:1}. \\

\noindent\textbf{Natural reconstruction} \cite{Barger:2003gt} \\
In the MSM, the Dirac matrix $M_{\rm D}$ of this parametrization including 6 arbitrary complex entries is expressed as
    \begin{eqnarray}
   M_{\rm D}= \begin{pmatrix}
        a_1&b_1\\
        a_2&b_2\\
        a_3&b_3
    \end{pmatrix} \,.
\end{eqnarray}
The approximate seesaw formula, $M_{\nu}\simeq -M_{\rm D}D_N^{-1}M_{\rm D}^{T}$, together with $\det(M_\nu) = 0$ in the MSM, provides $6-1$ identities.
If $M_1,M_2$ are given and any one entry of $M_{\rm D}$ is known, all the rest entries can be solved analytically. 
Without loss of generality, we assume $a_1 = |a_1| e^{i\arg a_1}$ is known. Then, the rest entries are solved to be
\begin{align}
	b_1&\simeq \zeta_1 \sqrt{-(M_{\nu})_{11}M_2-\frac{M_2}{M_1}a_1^2} \,, \nonumber\\
	a_\alpha &\simeq \frac{1}{(M_\nu)_{11}} \left\{ a_1\, (M_\nu)_{1\alpha} + \zeta_\alpha\, b_1\, \sqrt{\frac{M_1}{M_2}} \sqrt{(M_\nu)_{11} (M_\nu)_{\alpha\alpha} - (M_\nu)_{1\alpha}^2} \right\} \,, \quad \alpha=2,3, \nonumber\\
	b_\alpha &\simeq \frac{1}{(M_\nu)_{11}} \left\{ b_1\, (M_\nu)_{1\alpha} - \zeta_\alpha\, a_1\, \sqrt{\frac{M_2}{M_1}} \sqrt{(M_\nu)_{11} (M_\nu)_{\alpha\alpha} - (M_\nu)_{1\alpha}^2} \right\} \,, \quad \alpha=2,3.	
\end{align}
Here, $\alpha=1,2,3$ is a flavor index representing $e,\mu,\tau$.
Therefore, the 4 independent parameters from heavy neutrinos are $M_1$, $M_2$, $|a_1|$ and $\arg a_1$, as summarized in Table~\ref{tab:1}. One can replace $|a_1|$ and $\arg a_1$ by any $|a_i|$ and $\arg a_i$ or $|b_1|$ and $\arg b_1$. This parametrization is called the Natural Reconstruction in the review \cite{Guo:2006qa}. \\

\noindent\textbf{Blennow--Fernandez-Martinez (BFM) parametrization \cite{Blennow:2011vn}}\\
In MSM, the $5\times 5$ unitary matrix $\mathcal{U}$ who diagonalizes the neutrino mass matrix is parametrized as
\begin{eqnarray}
	{\cal U} = \begin{pmatrix} c & s \\ -s^\dag & \hat{c} \end{pmatrix} \begin{pmatrix} V_0 & 0 \\ 0 & I \end{pmatrix} \,,
\end{eqnarray}
where the nonphysical heavy-heavy mixing angle and phase have been omitted and $V_0$ is the standard PMNS (taking $\rho=0$).
The first matrix on the right side of the above equation is an unitary rotation matrix expressed as a series expansion of the matrix exponential,
\begin{eqnarray}
	\begin{pmatrix} c & s \\ -s^\dag & \hat{c} \end{pmatrix} \equiv \begin{pmatrix} 
		\displaystyle\sum_{n=0}^{\infty}\frac{(-\Theta \Theta^\dag)^n}{(2n)!} & 
		\displaystyle\sum_{n=0}^{\infty}\frac{(-\Theta \Theta^\dag)^n}{(2n+1)!} \Theta \\ 
		-\displaystyle\sum_{n=0}^{\infty}\frac{(-\Theta^\dag \Theta)^n}{(2n+1)!} \Theta^\dag & 
		\displaystyle\sum_{n=0}^{\infty}\frac{(-\Theta^\dag \Theta)^n}{(2n)!} \end{pmatrix} \,,
\end{eqnarray} 
where the $3\times 2$ matrix $\Theta \equiv V_L\theta V_R^{\dagger}$ is parametrized as
\begin{align}
	\Theta \equiv V_L \begin{pmatrix} 0 & 0 \\ \theta_y & 0 \\ 0 & \theta_z \end{pmatrix} V_R^\dagger \ \ \text{(NH)} \,, \quad \text{or} \quad \
	\Theta \equiv V_L \begin{pmatrix} \theta_x & 0 \\ 0 & \theta_y \\ 0 & 0 \end{pmatrix} V_R^{\dagger}  \ \ \text{(IH)} \,.
\end{align}
Here, $(V_L)_{3\times 3}$ and $(V_R)_{2\times 2}$ are unitary matrices written as
\begin{align}
	\begin{aligned}
		V_L &= 
		\begin{pmatrix}
			1 & 0 & 0 \\
			0 & e^{-i \alpha_2^L} & 0 \\
			0 & 0 & e^{-i \alpha_3^L}
		\end{pmatrix}
		\begin{pmatrix}
			c_1 c_3 & s_1 c_3 & s_3e^{-i \delta^L} \\
			-s_1 c_2-c_1 s_2 s_3e^{i \delta_L}  & +c_1 c_2-s_1 s_2 s_3 e^{i \delta^L} & s_2 c_3 \\
			+s_1 s_2-c_1 c_2 s_3  e^{i \delta^L} & - c_1 s_2-s_1 c_2 s_3 e^{i \delta^L} & c_2 c_3
		\end{pmatrix} \,,\\
		V_R & = 
		\begin{pmatrix}
			c_R & s_Re^{-i \delta^R} \\
			-s_Re^{i \delta^R} & c_R
		\end{pmatrix}\,,
	\end{aligned}
\end{align}
where $c_i\equiv \cos(\theta_{i}^L)$, $s_i\equiv \sin(\theta_{i}^L)$ ($i=1,2,3$), $c_R\equiv \cos(\theta_R)$ and $s_R\equiv \sin(\theta_R)$. 
The diagonalization of the neutrino mass matrix tells us that $M_{\rm D}$ can be parametrized as
\begin{align}
	M_{\rm D}&=-V_L^{*}\cos(\hat{\theta}_L)\left(V_L^{T}V_0^*D_\nu V_0^{\dagger}V_L^{}\right)\sin(\theta)V_R^{\dagger}+V_L^*\sin{(\theta)}\left(V_R^{T} D_N V_R^{}\right)\cos{(\hat{\theta}_R)}V_R^{\dagger} \,,
\end{align}
where $\hat{\theta}_L\equiv{\rm diag}\{0,\theta_y,\theta_z\}$ (NH) or ${\rm diag}\{\theta_x,\theta_y,0\}$ (IH), and $\hat{\theta}_R\equiv{\rm diag}\{\theta_y,\theta_z\}$ (NH) or ${\rm diag}\{\theta_x,\theta_y\}$ (IH).
Following the diagonalization procedure, the unitary $V_R$ and $V_L$ can be connected through
\begin{align}
	V_L^{T}(V_0^*D_{\nu}V_0^{\dagger})V_L =-\tan{\theta}(V_R^{T}D_N V_R)\tan{\theta}^T \,.\label{equ}
\end{align}
After given the introduced $\theta_y$ and $\theta_z(\theta_x)$, we can solve other parameters in $V_L$ and $V_R$
\footnote{
	We can define $X\equiv V_L^*\tan(\theta)V_R^{T}$ and then the above Eq.~\eqref{equ} is equal to $M_\nu^*\equiv V_0^*D_{\nu}V_0^\dagger=-XD_NX^{T}$.
	Here, if we regard $|X_{11}|$ and $\arg{X_{11}}$ as independent parameters, the rest entries of $(X)_{3\times 2}$ can be solved algebraically.
	Because the matrix $\theta$ satisfies $\lambda(X^\dagger X)=\lambda((\tan\theta)^\dagger \tan\theta)$, where $\lambda(A)$ means eigenvalues of $A$, we can obtain the eigenvalues of $X^\dagger X$,
	\begin{align}
		n_{\pm}^2=\frac{1}{2}\sum_{i,j}|X_{ij}|^2\pm \sqrt{(\frac{1}{2}\sum_{i,j}|X_{ij}|^2)^2-\sum_{i\neq j }|X_{i1}X_{j2}-X_{i2}X_{j1}|^2} \,, \label{equ:eig}
	\end{align}
	where $n_\pm^2=\tan{\theta}_y^2,\tan{\theta}_z^2$ for the NH or $\tan{\theta}_x^2,\tan{\theta}_y^2$ for the IH respectively, and $i=1,2,3$ and $j=1,2$.
	Combining \eqref{equ:eig} with $X=V_L^*\tan(\theta)V_R^{T}$, $V_R$ and $V_L$ can be solved through $XX^{\dagger}=V_L^{} \tan\theta (\tan\theta)^\dagger V_{L}^{\dagger}$ and $X^{\dagger}X=V_R^{} (\tan\theta)^\dagger \tan\theta V_{R}^{\dagger}$.
}.
If considering $D_N\gg M_{\rm D}$ and assuming $\theta$ is tiny, $M_{\rm D}$ can be approximated to $(V_L^{*}\theta V_R^{T})D_N$ and \eqref{equ} will return to the well-known approximate seesaw formula.
Therefore, in this parametrization, $M_{\rm D}$ is parametrized by 7 parameters in light neutrinos, 2 heavy masses $M_1$ and $M_2$, $\theta_{y}$ and $\theta_z(\theta_{x})$, as summarized in Table~\ref{tab:1}.\\

\noindent\textbf{Modified Casas--Ibarra--Ross parametrization} \cite{Ibarra:2005qi} \\
Based on $M_{\rm D}^{}$ parametrized as Eq.~\eqref{eq:CIR}, a Hermitian matrix $P$ is introduced,
\begin{equation}
P\equiv M_{\rm D}^{}M_{\rm D}^\dagger\simeq V_0\sqrt{D_{\nu}}\left(\Omega\, D_N\, \Omega^{\dagger}\right)\sqrt{D_{\nu}}\,V_0^{\dagger}  \,.
\end{equation}
In the MSM, due to $m_{\rm lightest}=0$, the orthogonal condition $(V_0^{\dagger}P)_{1\alpha}=0$ in the NH and $(V_0^{\dagger}P)_{3\alpha}=0$ in the IH ($\alpha=e,\mu,\tau$) can correlate the diagonal entries $P_{\alpha\alpha}$ with $P_{12},\ P_{13},\ P_{23}$. For example, in the NH scenario, there are
\begin{align}
	\begin{aligned}
		P_{11} &= -\frac{P_{12}^*(V_0^*)_{21} + P_{13}^*(V_0^*)_{31}}{(V_0^*)_{11}} \,,\\
		P_{22} &= -\frac{P_{12}(V_0^*)_{11} + P_{23}^*(V_0^*)_{31}}{(V_0^*)_{21}} \,, \\
		P_{33} &= -\frac{P_{13}(V_0^*)_{11} + P_{23}^{}(V_0^*)_{21}}{(V_0^*)_{31}} \,,
	\end{aligned}
\end{align}
and the 2 complex phases $\arg P_{13}$, $\arg P_{23}$, can be solved by $P_{12},|P_{13}|,|P_{23}|$,
\begin{align}
	\begin{aligned}
		\arg P_{13} &= \frac{-i \, \text{Im}[P_{12}^{} (V_0)_{21} (V_0^*)_{11}] + \xi \sqrt{|P_{13}^{}|^2 |(V_0)_{11}|^2 |(V_0)_{31}|^2 - [\text{Im}(P_{12}^{} (V_0)_{21} (V_0^*)_{11})]^2}}{|P_{13}^{}| (V_0)_{31} (V_0^*)_{11}}\,,\\   
		\arg P_{23} &= \frac{+i \, \text{Im}[P_{12}^{} (V_0)_{21} (V_0^*)_{11}] + \xi \sqrt{|P_{23}^{}|^2 |(V_0)_{21}|^2 |(V_0)_{31}|^2 - [\text{Im}(P_{12}^{} (V_0)_{21} (V_0^*)_{11})]^2}}{|P_{23}^{}| (V_0)_{31} (V_0^*)_{21}} \,,
	\end{aligned}
\end{align}
where $\xi=\pm 1$.
The analogous procedure can be taken in the IH. 
This parametrization includes the 7 parameters in light neutrinos and the 4 new parameters $|P_{12}|$, $|P_{13}|$, $|P_{23}|$ and $\arg P_{12}$, as summarized in Table~\ref{tab:1}.\\

\noindent\textbf{Bi-unitary parametrization} \cite{Endoh:2002wm} \\
The Dirac matrix $M_{\rm D}$ can be parametrized as
\begin{eqnarray}
     M_{\rm D} = V_L \begin{pmatrix} 0 & 0 \\ d_2 & 0 \\ 0 & d_3 \end{pmatrix} U_R  \ \ \text{(NH)} \ \quad \text{or} \quad \
     M_{\rm D} = V_L \begin{pmatrix} d_1 & 0 \\ 0 & d_2 \\ 0 & 0 \end{pmatrix} U_R  \ \ \text{(IH)}
\end{eqnarray}
through the 2 unitary matrix $V_L$ and $U_R$ expressed as 
\begin{align}
	\begin{aligned}
		V_L &= \begin{pmatrix}
			c_1 c_3 & s_1 c_3 & s_3 \\
			-c_1 s_2 s_3 - s_1 c_2 e^{-i \delta_L} & -s_1 s_2 s_3 + c_1 c_2 e^{-i \delta_L} & s_2 c_3 \\
			-c_1 c_2 s_3 + s_1 s_2 e^{-i \delta_L} & -s_1 c_2 s_3 - c_1 s_2 e^{-i \delta_L} & c_2 c_3
			\end{pmatrix}
			\begin{pmatrix}
			1 & 0 & 0 \\
			0 & e^{-i \gamma_L} & 0 \\
			0 & 0 & e^{+i \gamma_L}
			\end{pmatrix} \,, \\
		U_R & = \begin{pmatrix}
				c_R & s_R \\
				-s_R & c_R
				\end{pmatrix}
				\begin{pmatrix}
				e^{-i \gamma_R} & 0 \\
				0 & e^{+i \gamma_R}
				\end{pmatrix} \,,
	\end{aligned}
\end{align} 
where $c_i = \cos \theta_{i}$, $s_i = \sin \theta_{i}$, ($i=1, 2, 3$), $c_R = \cos \theta_{R}$ and $s_R = \sin \theta_{R}$.
This parametrization contains 4 rotation angles $\theta_1$, $\theta_2$, $\theta_3$, $\theta_R$, 3 CP violating phases $\delta_L$, $\gamma_L$, $\gamma_R$,  2 light masses in the mass eigenvalue equation of $M_{\rm D}^\dagger M_{\rm D}^{}$, and rest 2 parameters $d_1(d_3)$ and $d_2$, as summarized in Table~\ref{tab:1}.
It is emphasized that $V_L$ is not the PMNS matrix, which is approximately given by $V_0\simeq V_LK_R$. Here, $K_R$ is obtained by diagonalizing the following mass matrix, e.g. in the NH case,
\begin{eqnarray}
\begin{pmatrix} 0 & 0 \\ d_2 & 0 \\ 0 & d_3 \end{pmatrix} U_R D_N^{-1} U_R^{T} \begin{pmatrix} 0 & d_2 & 0 \\ 0 & 0 & d_3 \end{pmatrix} = K_R D_\nu K_R^{T} \,,
\end{eqnarray}
where the approximate seesaw formula have been applied.\\ 

\noindent\textbf{Vector Representation} \cite{Fujihara:2005pv} \\
The $M_{\rm D}$ can be parametrized as 
\begin{eqnarray}
M_{\rm D} =(m_{D1} \mathbf{u}_1, m_{D2} \mathbf{u}_2)\,,
\end{eqnarray}
where $m_{D1}$ and $m_{D2}$ are two real mass parameters and $\mathbf{u}_i = (u_{ei}, u_{\mu i}, u_{\tau i})^{T}$ (for $i=1,2$) are two unit vectors satisfying $\mathbf{u}_1^{\dagger}\mathbf{u}_1=\mathbf{u}_2^{\dagger} \mathbf{u}_2=1$.
Without loss of generality, $\mathbf{u}_1$ and $\mathbf{u}_2$ are taken as real and complex respectively. 
The 7 independent parameters in $M_{\rm D}$ under the normalized condition can be set as $u_{e1}, u_{\mu 1}, |u_{e2}|, |u_{\mu 2}|$ and three phases $\arg u_{e2}$, $ \arg u_{\mu 2}$, $\arg u_{\tau 2}$.
Then, $u_{\tau 1}$ and $ |u_{\tau 2}| $ can be solved through
\begin{align}
	\begin{aligned}
		u_{\tau 1} &=\xi_1\sqrt{1-u_{e 1}^2-u_{\mu 1}^2} \,,\\
		|u_{\tau 2}|&=\xi_2\sqrt{1-|u_{e 2}|^2-|u_{\mu 2}|^2} \,,
	\end{aligned}
\end{align}
where $\xi_{1,2}=\pm 1$. 
After applying the approximate seesaw formula, $M_\nu\simeq-M_{\rm D}D_N^{-1}M_{\rm D}^T$, $m_{D1}$ and $m_{D2}$ can be obtained via $\lambda(M_DD_N^{-1}M_D^TM_D^*D_N^{-1}M_D^\dagger)=\lambda(M_\nu M_\nu^\dagger)$, where $\lambda(A)$ means the eigenvalues of $A$ matrix, so there are
\begin{align}
	\begin{aligned}
		m_{D1,D2}^4 \simeq &M_{1,2}^2\left( \frac{n_-^2 + n_+^2}{2} - {\rm Re}\left[\mathbf{u}_1^{\dagger} \mathbf{u}_2\right]^2 \frac{n_- n_+}{1-|\mathbf{u}_1^{\dagger} \mathbf{u}_2|^2} \right) \\
        & \pm M_{1,2}^2\sqrt{ \left( \frac{n_-^2 + n_+^2}{2} - {\rm Re}\left[\mathbf{u}_1^{\dagger} \mathbf{u}_2\right]^2 \cdot \frac{n_- n_+}{1-|\mathbf{u}_1^{\dagger} \mathbf{u_2}|^2} \right)^2 - \left( \frac{n_- n_+}{1-|\mathbf{u}_1^{\dagger} \mathbf{u}_2|^2} \right)^2 } \ \ .
	\end{aligned}
\end{align}
Here, $n_\pm^2$ are eigenvalues of $M_\nu M_\nu^\dagger$, i.e. $n_-=m_2$,  $n_+=m_3$ in the NH and $n_-=m_1$, $n_+=m_2$ in the IH.
This parametrization contains 4 mass parameters $m_2$, $m_3(m_1)$, $M_1$, $M_2$ and 7 independent parameters in unit vectors $\mathbf{u}_1$ and $\mathbf{u}_2$, as summarized in Table~\ref{tab:1}.\\

\section{Related loop functions}\label{app:loopfunc}
The related functions about the 1-loop correction are listed below to analyse the cLFV processes in section~\ref{sec:mu_e_sector} and more details are shown in paper \cite{Alonso:2012ji}.
Here, $\mathcal{U}$ is the unitary matrix as Eq.~\eqref{eq:mathU}, $R$ is the active-sterile mixing matrix as Eq.~\eqref{eq:R}, $\alpha,\beta=e,\mu,\tau$ are flavor indices, $i,j=1,2,3,4,5$ represent all neutrinos, and $C_{ij}=\sum_{\alpha}\mathcal{U}^*_{\alpha i} \mathcal{U}_{\alpha j}$.
The related loop functions are 
\begin{align}
	G_\gamma(x) &= -\frac{x(2x^2 + 5x - 1)}{4(1-x)^3} - \frac{3x^3}{2(1-x)^4} \ln x \,, \label{eq:Ggamma}\\ 
	F_\gamma(x) &= \frac{x(7x^2 - x - 12)}{12(1-x)^3} - \frac{x^2(x^2 - 10x + 12)}{6(1-x)^4} \ln x \,, \nonumber \\
	F_Z(x) &= -\frac{5x}{2(1-x)} - \frac{5x^2}{2(1-x)^2} \ln x \,, \nonumber \\
	G_Z(x, y) &= -\frac{1}{2(x-y)} \left[ \frac{x^2(1-y)}{1-x} \ln x - \frac{y^2(1-x)}{1-y} \ln y \right] \,, \nonumber \\
	H_Z(x, y) &= \frac{\sqrt{xy}}{4(x-y)} \left[ \frac{x^2 - 4x}{1-x} \ln x - \frac{y^2 - 4y}{1-y} \ln y \right] \,, \nonumber  \\
	F_{\text{Box}}(x,y) &= \frac{1}{x-y} \Bigg\{ \left(4+ \frac{xy}{4} \right) \left[ \frac{1}{1-x} + \frac{x^{2}}{(1-x)^{2}}\ln x- \frac{1}{1-y} - \frac{y^{2}}{(1-y)^{2}}\ln y \right] \nonumber \\
	&\quad -2xy \left[ \frac{1}{1-x} + \frac{x}{(1-x)^{2}}\ln x - \frac{1}{1-y} - \frac{y}{(1-y)^{2}}\ln y \right] \Bigg\} \,, \nonumber \\ 
	F_{X\text{Box}}(x,y)  &= \frac{-1}{x-y} \Bigg\{ \left(1+ \frac{xy}{4} \right) \left[ \frac{1}{1-x} + \frac{x^{2}}{(1-x)^{2}}\ln x - \frac{1}{1-y} - \frac{y^{2}}{(1-y)^{2}}\ln y \right]  \nonumber\\
	&\quad -2xy \left[ \frac{1}{1-x} + \frac{x}{(1-x)^{2}}\ln x - \frac{1}{1-y} - \frac{y}{(1-y)^{2}}\ln y \right]\Bigg\} \,,\nonumber \\
	G_{\text{Box}}(x,y) &= \frac{-\sqrt{xy}}{x-y} \Bigg\{ \left(4+xy \right) \left[ \frac{1}{1-x} + \frac{x}{(1-x)^{2}}\ln x - \frac{1}{1-y} - \frac{y}{(1-y)^{2}}\ln y \right] \nonumber \\
	&\quad -2 \left[ \frac{1}{1-x} + \frac{x^{2}}{(1-x)^{2}}\ln x - \frac{1}{1-y} - \frac{y^{2}}{(1-y)^{2}}\ln y \right]\Bigg\} \,. \nonumber
\end{align}
In terms of these functions, the formal factors are 
\begin{align}
	G^{\alpha\beta}_{\gamma} &= \sum_{i} \mathcal{U}^{*}_{\alpha i} \mathcal{U}_{\beta i} G_{\gamma}(x_{i}) = \sum_{N_i} R^{*}_{\alpha N_i} R_{\beta N_i} G_{\gamma}(x_{N_i}) \,,   \label{eq:Ggammaalphabeta}\\
	F^{\alpha\beta}_{\gamma} &= \sum_{i} \mathcal{U}^{*}_{\alpha i} \mathcal{U}_{\beta i}^{} F_{\gamma}(x_{i}) = \sum_{N_i} R^{*}_{\alpha N_i} R_{\beta N_i}^{} F_{\gamma}(x_{N_i}) \,,\nonumber \\
	F^{\alpha\beta}_{Z} &= \sum_{i j} \mathcal{U}^{*}_{\alpha j} \mathcal{U}_{\beta i}^{} \left[ \delta_{i j} F_{Z}(x_{i}) + C_{i j} G_{Z}(x_{i}, x_{j}) + C^*_{i j} H_{Z}(x_{i}, x_{j}) \right] \nonumber \\
	&= \sum_{N_i N_j} R^{*}_{\alpha N_j} R_{\beta N_i} \Bigl\{ \delta_{N_i N_j} (F_{Z}(x_{N_i}) + 2G_Z(0, x_{N_i}))  \nonumber \\
	&  \quad + C_{N_i N_j} \left[G_{Z}(x_{N_i}, x_{N_j}) - G_{Z}(0,x_{N_i}) - G_{Z}(0,x_{N_j})\right] + C^*_{N_i N_j} H_{Z}(x_{N_i},x_{N_j}) \Bigr\} \,, \nonumber \\
    F^{\alpha\beta\beta\beta}_{\text{Box}} &= \sum_{i j} \mathcal{U}^{*}_{\alpha j} \mathcal{U}_{\beta i} \left[ \mathcal{U}^*_{\beta j} \mathcal{U}_{\beta i} G_{\text{Box}}(x_{i},x_{j}) - 2 \mathcal{U}_{\beta j}\mathcal{U}^*_{\beta i} F_{X\text{Box}}(x_{i},x_{j})  \right] \nonumber \\
	&= -2 \sum_{N_i} R_{\alpha N_i}^{*} R_{\beta N_i} \left[ F_{X\text{Box}}(x_{N_i},0)-F_{X\text{Box}}(0,0) \right] \nonumber \\
	&\quad + \sum_{N_i N_j} R_{\alpha N_j}^{*} R_{\beta N_i} \Bigl\{ R_{\beta N_j}^{*} R_{\beta N_i} G_{\text{Box}}(x_{N_i},x_{N_j}) - 2 R_{\beta N_j} R_{\beta N_i}^{*} \bigl[ F_{X\text{Box}}(x_{N_i},x_{N_j}) \nonumber \\
	&\quad - F_{X\text{Box}}(0,x_{N_j}) - F_{X\text{Box}}(x_{N_i},0) + F_{X\text{Box}}(0,0) \bigr] \Bigr\} \,, \nonumber \\
	F_{\frac{1}{2}\text{Box}+Z}^{\alpha \beta\beta\beta} &= \frac{1}{2} F_{\text{Box}}^{\alpha \beta\beta\beta} + F_Z^{\alpha \beta}, \quad \quad {\rm and} \quad \quad F_{Z-\gamma}^{\alpha \beta}= 2 s_{\rm w}^2 \left( F_Z^{\alpha \beta} - F_\gamma^{\alpha \beta} \right) \,,   \label{eq:F}
\end{align}
where $x_i=m_i^2/M_W^2$ ($m_4=M_1,m_5=M_2$) and $x_{N_i}=M_{N_i}^2/m_W^2$.
The functions about quarks are
\begin{align}
	F_u(x) &= \frac{2}{3}s_{\rm w}^2 G_\gamma(x) + \frac{2}{3}s_{\rm w}^2 \left[ F_\gamma(x) - F_Z(x) - 2G_Z(0, x) \right] \nonumber \\
	&\quad + \frac{1}{4} \left[ F_Z(x) + 2G_Z(0, x) + F_{\text{Box}}(x, 0) - F_{\text{Box}}(0, 0) \right] \,, \nonumber \\
	F_d(x) &= - \frac{1}{3}s_{\rm w}^2 G_\gamma(x) - \frac{1}{3}s_{\rm w}^2 \left[ F_\gamma(x) - F_Z(x) - 2G_Z(0, x) \right] \nonumber \\
	&\quad - \frac{1}{4} \left[ F_Z(x) + 2G_Z(0, x) - F_{X\text{Box}}(x, 0) + F_{X\text{Box}}(0, 0) \right] \,. \label{eq:Fud}
\end{align}
When $x\gg 1$, $F_u(x)$ and $F_d(x)$ can be simplified as 
\begin{align}
	F_u(x)& \underset{x\gg 1}{\longrightarrow} \frac{2}{3}s_{\rm w}^2\frac{16\ln(x)-31}{12}-\frac{3+3\ln(x)}{8} \,, \nonumber \\
	F_d(x)& \underset{x\gg 1}{\longrightarrow} -\frac{1}{3}s_{\rm w}^2\frac{16\ln(x)-31}{12}-\frac{3-3\ln(x)}{8} \,. \label{eq:Fudsim}
\end{align}

\end{document}